\documentclass[prd, twocolumn, amsfonts, amssymb, amsmath, showkeys, nofootinbib, superscriptaddress, floatfix, longbibliography]{revtex4-2}

\usepackage[english]{babel}
\usepackage[utf8]{inputenc}
\usepackage{amsthm}
\usepackage{mathtools}
\usepackage{physics}
\usepackage{graphicx}
\usepackage{dcolumn}
\usepackage{bm}
\usepackage{xcolor}
\usepackage[pdftex, pdftitle={Article}, pdfauthor={Author}]{hyperref}
\usepackage{multirow}
\usepackage{hhline}
\usepackage{aas_macros}
\usepackage{orcidlink}
\usepackage[normalem]{ulem}

\newcommand{\msun}{\mathrm{\, M_\odot}}
\newcolumntype{P}[1]{>{\centering\arraybackslash}p{#1}}

\begin{document}

\preprint{APS/123-QED}

\title{Testing general relativity with amplitudes of subdominant gravitational-wave modes}

\author{Ish Gupta\orcidlink{0000-0001-6932-8715}}
\email[Correspondence email address: ]{ishgupta@berkeley.edu}
\affiliation{Department of Physics, University of California, Berkeley, CA 94720, USA}
\affiliation{Department of Physics and Astronomy, Northwestern University, 2145 Sheridan Road, Evanston, IL 60208, USA}
\affiliation{Center for Interdisciplinary Exploration and Research in Astrophysics (CIERA), Northwestern University, 1800 Sherman Ave, Evanston, IL 60201, USA}
\affiliation{Institute for Gravitation and the Cosmos, Department of Physics, Pennsylvania State University, University Park, PA 16802, USA}

\author{Purnima Narayan\orcidlink{0009-0009-0599-532X}}
\affiliation{Department of Physics and Astronomy, The University of Mississippi, University, Mississippi 38677, USA}

\author{Lionel London\orcidlink{0000-0001-8239-4370}}
\affiliation{King’s College London, Strand, London WC2R 2LS, United Kingdom}

\author{Shubhanshu Tiwari\orcidlink{0000-0003-1611-6625}}
\affiliation{Physics Institute, University of Z\"{u}rich, 8057, Switzerland}

\author{Bangalore Sathyaprakash\orcidlink{0000-0003-3845-7586}}
\affiliation{Institute for Gravitation and the Cosmos, Department of Physics, Pennsylvania State University, University Park, PA 16802, USA}

\begin{abstract}
We present an improved subdominant-mode amplitude (SMA) test of general relativity (GR), which probes amplitude-level deviations in the higher-order modes of gravitational-wave (GW) signals from binary black hole mergers while keeping the dominant quadrupole mode fixed. Using a comprehensive parameter-estimation campaign, we benchmark the test against Gaussian noise fluctuations, waveform modeling systematics, and physical effects such as spin precession and orbital eccentricity. When applied to numerical-relativity simulations, the SMA test performs reliably for aligned-spin and mildly precessing systems but exhibits measurable biases for strongly precessing or eccentric binaries. Although designed to detect amplitude deviations, the test also responds coherently to phase perturbations, yielding apparent GR violations when applied to phase-modified waveforms. Applied to recent GW detections, we report the strongest constraint on the hexadecapolar $(4,4)$ mode amplitude deviation, $\delta A_{44} = -0.30^{+1.16}_{-3.45}$, consistent with GR. With these results, this work establishes the SMA test as a robust and broadly sensitive null test of general relativity and demonstrates a systematic approach for assessing the robustness of GW tests of GR.
\end{abstract}

\maketitle

\section{Introduction} \label{sec:intro}

The direct detection of gravitational waves (GWs) from compact binary mergers by the LIGO-Virgo-KAGRA (LVK) \cite{KAGRA:2013rdx,LIGOScientific:2014pky,VIRGO:2014yos,KAGRA:2020tym,Somiya:2011np,Virgo:2019juy,aLIGO:2020wna,Capote:2024rmo} observatories has transformed experimental gravity into a precision science. These signals provide access to regimes of extreme curvature and strong-field dynamics that were previously inaccessible, allowing general relativity (GR) to be tested under the most demanding conditions known in nature. The first detection, GW150914~\cite{LIGOScientific:2016aoc}, was confirmed to be consistent with the predictions of GR~\cite{LIGOScientific:2016lio}, and the growing catalog of more than 150 binary black hole (BBH) mergers up to GWTC-4~\cite{LIGOScientific:2025slb} now constrains potential deviations from GR across the inspiral, merger, and ringdown regimes~\cite{LIGOScientific:2021sio,LIGOScientific:2026qni,LIGOScientific:2026fcf,LIGOScientific:2026wpt}.  

The precision of these tests depends critically on the accuracy of waveform models that describe the complex dynamics of binary coalescence. The GW waveform from a BBH system can be decomposed into multipolar components,
\begin{equation}
    h(t,\boldsymbol{\lambda}) = \sum_{\ell=2}^{\infty}\sum_{m=-\ell}^{\ell}
    Y_{-2}^{\ell m}(\theta_{JN},\phi_0)\,h_{\ell m}(t,\boldsymbol{\lambda}),
    \label{eq:ht}
\end{equation}
where $Y_{-2}^{\ell m}$ are the spin-weighted spherical harmonics of weight $-2$, $h_{\ell m}$ are the individual mode amplitudes that depend on the binary parameters $\boldsymbol{\lambda}$, $\theta_{JN}$ refers to the angle between the total angular momentum and the line of sight, and $\phi_0 = 0$ by convention~\cite{Pratten:2020ceb}. While the quadrupolar $(\ell,m)=(2,\pm2)$\footnote{Henceforth, we will represent $(\ell,\pm m)$ modes as $(\ell,m)$, unless stated otherwise.} modes dominate most observed signals, higher-order modes (HOMs) become significant in systems with unequal masses, large total mass, or precessing spins~\cite{Chatziioannou:2014bma,Roy:2019phx,Gupta:2024bqn}. These HOMs encode complementary information about the geometry and dynamics of spacetime near merger, making them powerful and largely untapped probes of the theory of gravity itself.

Parameterized tests of GR typically introduce fractional deviations in the phase evolution of the waveform, most often through post-Newtonian (PN) phasing coefficients~\cite{Agathos:2013upa,LIGOScientific:2016lio,LIGOScientific:2018dkp,Mehta:2022pcn,LIGOScientific:2026fcf}. Here, we pursue a complementary route that probes the amplitude structure of the signal. The \textit{subdominant-mode amplitude} (SMA) test~\cite{Islam:2019dmk,Puecher:2022sfm} modifies the amplitudes of the subdominant modes while keeping the dominant quadrupole mode fixed:
\begin{align}
h(t; \theta_{JN}, \boldsymbol{\lambda})
&= \sum_{m=\pm2} Y^{-2}_{2m}(\theta_{JN},0)\,h_{2m}(t;\boldsymbol{\lambda}) \notag\\
&\quad + \sum_{\text{HOM}} (1+\delta A_{\ell m})\,Y^{-2}_{\ell m}(\theta_{JN},0)\,h_{\ell m}(t;\boldsymbol{\lambda}),
\end{align}
where $\delta A_{\ell m}$ quantifies potential departures from the GR-predicted amplitude. While modified theories of gravity may induce more general deformations of the waveform amplitude and phase~\citep{Mezzasoma:2022pjb}, the SMA framework is intended here as a phenomenological consistency test of GR rather than as a direct mapping to a particular non-GR theory. In this sense, it probes whether the amplitude of a given subdominant $(\ell,m)$ mode is consistent with the GR prediction once the dominant quadrupole mode is held fixed. \citet{Puecher:2022sfm} implemented the test by allowing deviations to the $(\ell,m)=\{(2,1), (3,3)\}$ modes. When applied to GW190412~\cite{LIGOScientific:2020stg} and GW190814~\cite{LIGOScientific:2020zkf}, two mass-asymmetric events from GWTC-3~\cite{KAGRA:2021vkt}, the test resulted in informative bounds on $\delta A_{21}$ and $\delta A_{33}$ consistent with GR~\cite{LIGOScientific:2026qni}.

In this work, we extend the SMA formalism to include the $(4,4)$ and $(3,2)$ modes and evaluate its performance across a broad range of BBH configurations. Unlike $(2,1)$ and $(3,3)$, both the $(3,2)$ and the $(4,4)$ modes contribute to the signal for equal mass binaries (see Sec.~\ref{sec:homs} for more details). Furthermore, unlike other HOMs, $(3,2)$ contributes even for face-on orientations, thereby extending the reach of amplitude-based tests to the majority of detectable systems. We implement the test using the IMRPhenomXPHM waveform model~\cite{Pratten:2020ceb,Colleoni:2024knd}, which incorporates the dominant $(2,2)$ mode and the sub-dominant $(3,3)$, $(4,4)$, and $(3,2)$ modes, while also including the effect of spin-orbit precession.

We find that the SMA test reliably recovers injected amplitude deviations and produces null results for GR-consistent signals, validating its statistical behavior. When applied to numerical-relativity simulations, the test remains robust for aligned or mildly precessing systems but shows measurable bias when waveform systematics or missing physical effects, such as strong spin-orbit precession or orbital eccentricity, become significant. Beyond its immediate role as a null test for amplitude consistency, we demonstrate that the SMA framework can also capture deviations in GW phase. Simulated signals with phase deviations at specific PN orders yield non-zero estimates of $\delta A_{\ell m}$, showing that the test 
may be sensitive to a broad class of beyond-GR effects, even those that formally enter through the waveform phase. 
We further apply the method to recent detections, including GW230814~\cite{2025arXiv250907348T}, and obtain the strongest constraint on the amplitude deviation of the hexadecapolar $(4,4)$ mode till date.  

Equally important, this work establishes a rigorous standard for evaluating such null tests of GR. We systematically benchmark the SMA framework through Gaussian-noise realizations, injection–recovery studies, and comparisons with numerical-relativity simulations. We further apply the method to real GW detections and analyze parameter degeneracies that can mimic non-GR behavior. Together, these steps demonstrate both the reliability of the SMA test and provide a general template to quantify the robustness of tests of GR with GWs. Taken together, our results position the SMA test as a conceptually distinct and experimentally validated null test of GR. The remainder of this paper is organized as follows: Sec.~\ref{sec:homs} discusses the contribution of HOMs to the waveform; Sec.~\ref{sec:bench} benchmarks the SMA test against noise and numerical-relativity simulations; Sec.~\ref{sec:non-gr} explores its response to simulated deviations from GR; and Sec.~\ref{sec:gw_events} presents results when applying the test to a subset of GW events observed in LVK's fourth observing run (O4). Section~\ref{sec:concl} summarizes our conclusions.


\section{Contribution of Higher-Order Modes} \label{sec:homs}

\begin{figure}[!htbp]
    \centering
    \includegraphics[width=0.99\linewidth]{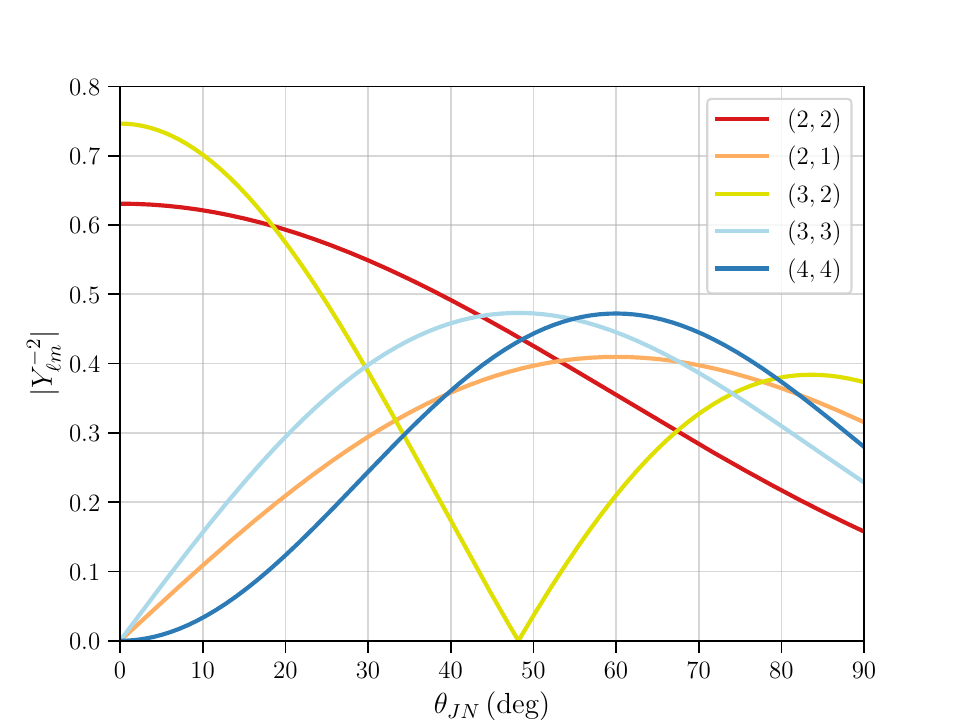}
    \caption{Absolute value of $Y^{-2}_{\ell m}$ as a function of $\theta_{JN}$ for different $(\ell,m)$ modes. Among the listed modes, only $(2,2)$ and $(3,2)$ contribute for face-on binaries, and other HOMs become relatively more important as inclination increases.}
    \label{fig:thetajn_ylm}
\end{figure}

The constraints on $\delta A_{\ell m}$ depend on the signal-to-noise ratio (SNR) of the corresponding $(\ell, m)$ mode, which in turn is strongly influenced by the binary’s inclination $\theta_{JN}$. This dependence arises through $Y^{-2}_{\ell m}(\theta_{JN},0)$. Figure~\ref{fig:thetajn_ylm} illustrates the absolute value of $Y^{-2}_{\ell m}$ as a function of $\theta_{JN}$ for various modes. Notably, for face-on binaries ($\theta_{JN} = 0$), only the $(2,2)$ and $(3,2)$ modes contribute significantly. As $\theta_{JN}$ increases, the relative contribution of HOMs grows, while that of the dominant quadrupole mode diminishes.

\begin{figure*}[!htbp]
    \centering
    \includegraphics[trim={0.4cm 0.4cm 0.4cm 0.3cm},width=0.9\linewidth]{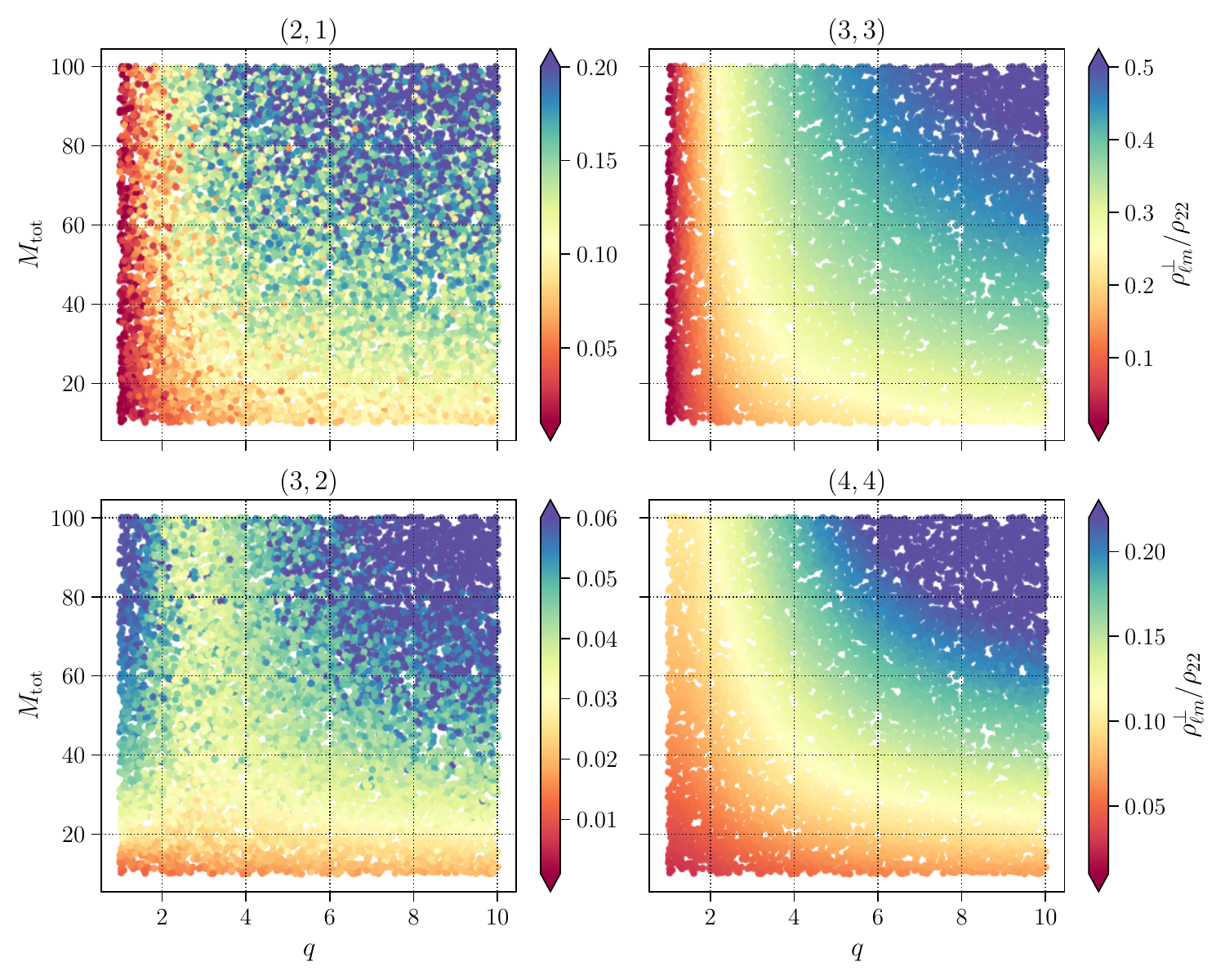}
    \caption{The ratio of the orthogonal , $\rho_{\ell m}^{\perp}$, in the $(\ell,m)$ modes compared to the $(2,2)$ mode, as a function of the total mass $M_{\rm tot}$ and the mass ratio $q$ of the binary.}
    \label{fig:rho_lm}
\end{figure*}

Beyond the viewing angle, the detectability of HOMs also depends on the binary's intrinsic parameters. In the inspiral regime, the phase of a given $(\ell, m)$ mode is related to the quadrupolar phase by 
\begin{equation}
    \phi_{\ell m} = \frac{m}{2} \phi_{22},
\end{equation}
leading to merger frequencies of $1.5 f_{\rm merge}$ and $2 f_{\rm merge}$ for the $(3,3)$ and $(4,4)$ modes, respectively. Since the sensitivity of current LVK detectors peaks in the $\mathcal{O}(100)$ Hz range, increasing the total binary mass $M_{\rm tot}$ lowers $f_{\rm merge}$ and shifts the HOMs into the sensitive band, thereby enhancing their contribution to the signal.

Focusing on the GW amplitude, the dominant $(2,2)$ mode appears at leading order in the PN expansion, corresponding to 0PN. In contrast, the $(2,1)$ and $(3,3)$ modes enter at 0.5PN, while the $(3,2)$ and $(4,4)$ modes appear at 1PN \cite{Pan:2010hz,Blanchet:2013haa,Gupta:2024bqn}. In terms of the PN expansion coefficient, given by
\begin{equation}
    v = (\pi M_{\rm tot} f_{22})^{1/3},
\end{equation}
where $f_{22}$ is the frequency of the $(2,2)$ mode. HOMs become increasingly significant at larger $v$. This dependence enhances the contribution of HOMs in systems with higher masses. Furthermore, the amplitudes of odd-$m$ modes depend on $q$ \cite{Pan:2010hz,Blanchet:2013haa}. These modes vanish in the equal-mass limit and grow in significance for asymmetric mass configurations.

To quantify the impact of intrinsic parameters on HOM significance, we compute the relative importance of an HOM with respect to the $(2,2)$ mode for a population of BBH systems with spins aligned to the orbital angular momentum. Specifically, we evaluate the SNR of the mode component orthogonal to the $(2,2)$ mode, $\rho^{\perp}_{\ell m}$, which measures the fraction of the signal that cannot be attributed to the dominant quadrupole mode \cite{Mills:2020thr}. This is given by,
\begin{equation}
    \rho_{\ell m}^{\perp} = \sqrt{(\tilde{h}_{\ell m},\tilde{h}_{\ell m})- \frac{(\tilde{h}_{\ell m},\tilde{h}_{22})^{2}}{(\tilde{h}_{22},\tilde{h}_{22})}},
\end{equation}
where $\tilde{h}_{\ell m}$ is the frequency-domain $(\ell, m)$ mode waveform, and $(\tilde{h}_{\ell m},\tilde{h}_{22})$ denotes the noise-weighted scalar product between the $(\ell, m)$ mode and the $(2,2)$ mode~\cite{Mills:2020thr,Hoy:2020vys}.

Figure~\ref{fig:rho_lm} presents the relative strength of different HOMs with respect to the $(2,2)$ mode as a function of $M_{\rm tot}$ and $q$. As expected, the contribution of HOMs increases with both $M_{\rm tot}$ and mass asymmetry. While the $(2,2)$ mode remains dominant across the parameter space considered, the $(3,3)$ mode consistently emerges as the second-strongest for asymmetric binaries, followed by $(2,1)$ and $(4,4)$~\citep{Mills:2020thr,Pitte:2023ltw,Yi:2025pxe,Yi:2026ucv}. The $(3,2)$ mode contributes the least, primarily due to its significant overlap with the $(2,2)$ mode, quantified by $(\tilde{h}_{32},\tilde{h}_{22})$, which suppresses the orthogonal SNR component, $\rho_{32}^{\perp}$. Nevertheless, for nearly equal-mass systems, both the $(3,2)$ and $(4,4)$ modes can become relevant, particularly when the total SNR is sufficiently high.

While the analysis here is restricted to BBHs with spins aligned to the orbital angular momentum, it is important to note that spin-orbit misalignment, leading to spin precession, can further enhance HOM contributions \cite{Chatziioannou:2014bma}. In particular, the modes in the inertial frame, $h^{J}_{\ell m}$, which track the total angular momentum $\boldsymbol{J}$, are expressed as linear combinations of modes in the co-precessing frame, $h^{L}_{\ell m'}$, which track the orbital angular momentum $\boldsymbol{L}$. This mixing occurs over $m' \in [-\ell, \ell]$, leading to an effective redistribution of power among modes~\cite{Pratten:2020ceb}. As a result, precession-induced mixing can amplify certain HOMs in the inertial frame, such as the $(2,1)$ mode due to contributions from the co-precessing $(2,2)$ mode, and the $(3,2)$ mode arising from the co-precessing $(3,3)$ mode. While this amplification makes mass-asymmetric, highly precessing binaries ideal candidates for the SMA test, it also places them in a regime where waveform systematics are particularly pronounced~\cite{Dhani:2024jja}. The impact of these systematics on the SMA test for significantly precessing systems will be explored in Sec.~\ref{sec:bench}.


\section{Benchmarking the test} \label{sec:bench}

For a parametric test of GR such as SMA, it is crucial to identify factors that can bias the estimation of deviation parameters, potentially leading to a false violation of GR~\citep{Saini:2022igm,Narayan:2023vhm,Chandramouli:2024vhw}. Such biases can arise due to both Gaussian and non-stationary noise, unmodeled physical effects such as interactions with a binary’s environment or orbital eccentricity, and systematic waveform modelling inaccuracies, among other sources (see~\citet{Gupta:2024gun} for a review). In this section, we assess the robustness of the SMA test against some of these effects.

Section~\ref{subsec:bayes} outlines the Bayesian formalism for inferring probability distributions on the deviation parameters and quantifying potential violations of GR. In Sec.~\ref{subsec:gauss_noise}, we examine the impact of Gaussian noise on the inference of $\delta A_{\ell m}$. In Sec.~\ref{subsec:NR}, we investigate whether waveform systematics can lead to spurious deviations in $\delta A_{\ell m}$ by applying the SMA test to numerical relativity simulations from the Simulating eXtreme Spacetimes (SXS) catalog~\cite{Boyle:2019kee}. Additionally, we assess the SMA test’s performance on SXS simulations incorporating significant orbital eccentricity to determine whether the recovered $\delta A_{\ell m}$ posteriors reject the GR hypothesis due to missing physical effects in the waveform model.


\subsection{Bayesian inference formalism} \label{subsec:bayes}

The probability distribution for $\delta A_{\ell m}$ is obtained via Bayesian parameter estimation on GW data. The detector response, $d$, is modeled as
\begin{equation}
    d(t,\boldsymbol{\theta}) = n(t) + s(t,\boldsymbol{\theta}),
\end{equation}
where $n(t)$ is the noise and $s(t,\boldsymbol{\theta})$ is the GW signal, which depends on the binary parameters $\boldsymbol{\lambda}$ and the source’s sky location, all collectively represented by $\boldsymbol{\theta}$. The posterior distribution $p(\boldsymbol{\theta}|d)$ is given by Bayes' theorem:
\begin{equation}
    p(\boldsymbol{\theta}|d) = \frac{p(d|\boldsymbol{\theta})\,\pi(\boldsymbol{\theta})}{\mathcal{Z}},
\end{equation}
where $p(d|\boldsymbol{\theta})$ is the likelihood function, $\pi(\boldsymbol{\theta})$ is the prior on the parameters, and $\mathcal{Z}$ is the Bayesian evidence,
\begin{equation}
    \mathcal{Z} = \int p(d|\boldsymbol{\theta})\,\pi(\boldsymbol{\theta})\,d\boldsymbol{\theta}.
\end{equation}
The marginalized posterior distribution for $\delta A_{\ell m}$ is then obtained as
\begin{equation}
    p(\delta A_{\ell m}|d) = \int \left( \prod \limits_{\theta_i \neq \delta A_{\ell m}} d\theta_i\right)\,p(\boldsymbol{\theta}|d).
\end{equation}
A deviation from GR is identified if $\delta A_{\ell m} = 0$ lies outside the $99\%$ credible interval (CI), corresponding to a $\sim3\sigma$ significance.

An alternative method for quantifying deviations from GR is Bayesian model selection, characterized by the logarithm of the Bayes factor, $\log \mathcal{B}_{\rm GR}^{\rm non-GR}$:
\begin{equation}
    \log \mathcal{B}_{\rm GR}^{\rm non-GR} = \log \frac{\mathcal{Z}^{\rm non-GR}}{\mathcal{Z}^{\rm GR}},
\end{equation}
where $\mathcal{Z}^{\rm GR}$ is the evidence for the hypothesis that the signal is consistent with GR, and $\mathcal{Z}^{\rm non-GR}$ is the evidence for an apparent deviation from GR. The evidence $\mathcal{Z}^{\rm GR}$ is obtained by performing parameter estimation while fixing $\delta A_{\ell m} = 0$ for all HOMs, while $\mathcal{Z}^{\rm non-GR}$ is computed by allowing $\delta A_{\ell m}$ to vary for a given $(\ell, m)$ mode. A value of $\log \mathcal{B}_{\rm GR}^{\rm non-GR} \gtrsim 5$ constitutes strong evidence for an apparent violation of GR.

It is important to note that a low posterior support for $\delta A_{\ell m} = 0$ does not necessarily imply a large value of $\log \mathcal{B}_{\rm GR}^{\rm non-GR}$. This discrepancy arises because the evidence calculation for the \textit{non-GR} model involves integrating the likelihood $p(d|\boldsymbol{\theta})$ over a larger prior space, due to the additional $\delta A_{\ell m}$ parameter, compared to the GR model. This may result in a lower $\mathcal{Z}^{\rm non-GR}$ due to Occam's penalty, which disfavors models with higher complexity unless they provide a significantly better fit to the data. Moreover, since Bayes factors are intrinsically prior-dependent, their numerical value can depend sensitively on the adopted prior range for $\delta A_{\ell m}$. Even if the likelihood is well-localized within the prior boundaries, choosing arbitrarily wide priors decreases the normalized prior density and can reduce $\mathcal{Z}^{\rm non-GR}$, and hence the corresponding Bayes factor, through an additional Occam penalty. This should be kept in mind when interpreting Bayes factors for hypothesis testing. In subsequent sections, we report $\log \mathcal{B}_{\rm GR}^{\rm non-GR}$ (abbreviated to $\log \mathcal{B}$ for brevity) for all cases where the $\delta A_{\ell m}$ posteriors indicate a significant deviation from 0.


\subsection{Effect of Gaussian noise} \label{subsec:gauss_noise}

\begin{figure}[!htbp]
    \centering
    \includegraphics[trim={0.4cm 0.4cm 0.4cm 0.3cm},width=0.99\linewidth]{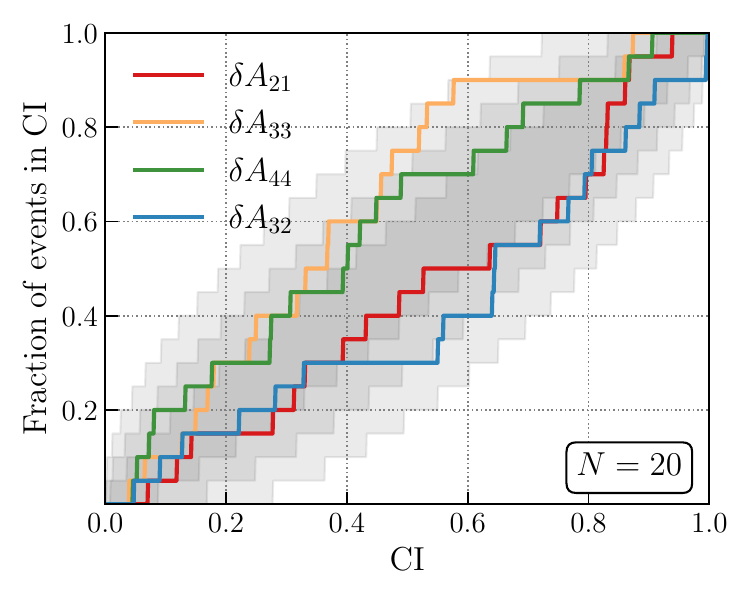}
    \caption{P-P plot for the deviation parameters in a $(32\msun,8\msun)$ BBH across 20 Gaussian noise realizations. The shaded regions indicate the $1\sigma$, $2\sigma$, and $3\sigma$ confidence intervals (CI).}
    \label{fig:pp_plot}
\end{figure}

\begin{figure*}[!htbp]
    \centering
    \includegraphics[trim={0.3cm 0.4cm 0.4cm 0.3cm},clip,width=0.8\linewidth]{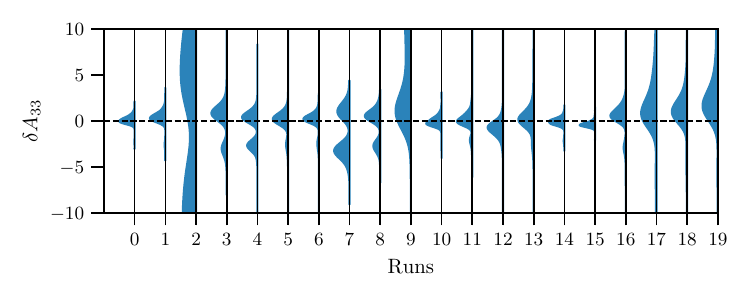}
    \caption{Probability distributions for $\delta A_{33}$ across 20 parameter estimation runs in different Gaussian noise realizations. Several runs show significant bimodality in the $\delta A_{33}$ posteriors, which is attributed to the effect of Gaussian noise and parameter degeneracies (see Appendix~\ref{app:degeneracy} for more details).}
    \label{fig:gauss_c33_violin}
\end{figure*}

To evaluate the impact of Gaussian noise on the estimation of $\delta A_{\ell m}$, we simulate a moderately precessing $(48\msun,12\msun)$ binary with an inclination angle $\theta_{JN} \sim 50^{\circ}$ and an effective precession parameter \cite{Schmidt:2014iyl} of $\chi_p = 0.2$. The GW signal is generated using IMRPhenomXPHM with $\delta A_{\ell m} = 0$ for all HOMs and injected into 20 independent Gaussian noise realizations consistent with Advanced LIGO and Virgo sensitivities. For parameter estimation, we allow $\delta A_{\ell m}$ to vary within a uniform prior of $[-10,10]$, considering one HOM at a time. The analysis is performed using the dynesty sampler \cite{Speagle:2019ivv} within the Bilby framework \cite{Ashton:2018jfp}. For each $(\ell, m)$ mode, the luminosity distance $D_L$ for the corresponding runs is set such that the optimal orthogonal SNR, $\rho^{\perp}_{\ell m}$, is approximately 3, ensuring that the SMA test for the mode is statistically meaningful.

Under ideal statistical behavior, the injected value $\delta A_{\ell m} = 0$ should lie within the $N\%$ confidence interval (CI) for $N\%$ of the cases. To assess this, we present the percent-percent (p-p) plot in Fig.~\ref{fig:pp_plot}, where the shaded regions represent $1\sigma$, $2\sigma$, and $3\sigma$ confidence intervals. The p-p curves for all HOMs remain within the $3\sigma$ contour, indicating consistency with expectations. However, we observe a notable deviation for $\delta A_{33}$, which reaches the $3\sigma$ boundary. Specifically, for more than 80\% of cases, the value $\delta A_{33} = 0$ falls within the $60\%$ CI, deviating from statistical expectations.

To investigate this, we examine the $\delta A_{33}$ posteriors from individual runs, shown in Fig.~\ref{fig:gauss_c33_violin}. In several noise realizations, the posterior distributions exhibit bimodal features, with one peak near the injected value $\delta A_{33} = 0$ and another displaced away from it. \citet{Puecher:2022sfm} posits that this bimodality arises due to a correlation between $\iota$ and $\delta A_{33}$. We find that the bimodality is further exacerbated by the degeneracy between $\delta A_{33}$ and the reference phase (see Appendix~\ref{app:degeneracy} for details). The presence of multiple modes broadens the CIs, leading to a higher-than-expected fraction of cases where $\delta A_{33} = 0$ lies within the $60\%$ CI.

While such tests are suitable for a larger set of parameter estimation runs, we restrict ourselves to $N=20$ due to computational cost. However, this restricted set already points to non-trivial features in the $\delta A_{\ell m}$ recovery due to Gaussian noise effects and parameter degeneracies.


\subsection{Inference with numerical relativity simulations} \label{subsec:NR}
As the SMA test is implemented using the IMRPhenomXPHM waveform, it is essential to assess whether waveform systematics or missing physical effects could bias the estimation of deviation parameters. To investigate this, we inject numerical relativity simulations from the SXS catalog into zero noise, corresponding to LIGO/Virgo sensitivities in O4~\cite{Capote:2024rmo}, and perform the SMA test on these simulated signals. In Sec.~\ref{subsubsec:wave_sys}, we analyze quasi-spherical BBH simulations with varying source-frame total mass, mass ratios, and in-plane spin components, leading to different levels of spin-orbit precession. In Sec.~\ref{subsubsec:ecc}, we examine two eccentric BBH simulations to determine whether unmodeled eccentricity features can mimic violations of GR.

\subsubsection{Effect of waveform systematics} \label{subsubsec:wave_sys}
To assess waveform modeling–induced biases, we apply the SMA test to nine SXS simulations, summarized in Table~\ref{tab:nr_runs}. Because HOM content depends on the total mass $M_{\rm tot}$ and mass ratio $q$, we consider $M_{\rm tot}\in\{100,200\}\msun$ and $q\in[1.1,7]$. Spin–orbit precession further redistributes power among modes and high precession is a regime where present models are least accurate~\cite{Dhani:2024jja}. We therefore include systems with substantial in–plane spin on the primary, $|\boldsymbol{\chi}_{1,\perp}|\in[0.27,0.89]$. For each simulation, the inclination angle $\iota = 60^{\circ}$, and the luminosity distance $D_L$ is set so that the three–detector LIGO Hanford, LIGO Livingston, and Virgo network attains an optimal SNR of 50.

\begin{figure*}[!htbp]
    \centering
    \includegraphics[trim={0.4cm 0.4cm 0.4cm 0.3cm},width=0.99\linewidth]{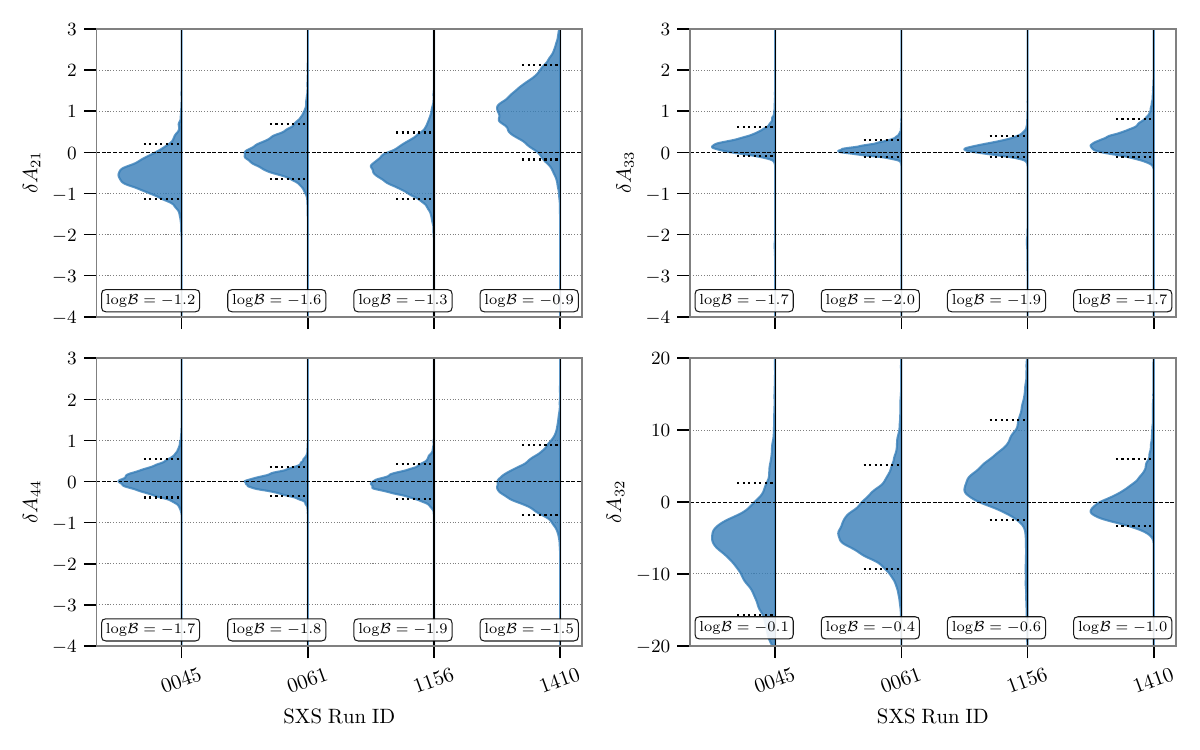}
    \caption{Posterior distributions for $\delta A_{\ell m}$ from the SMA test on SXS simulations listed in Tab.~\ref{tab:nr_runs}. The black dashed line marks $\delta A_{\ell m} = 0$, while the dotted lines denote the $90\%$ credible intervals. $\log \mathcal{B}$ represents the logarithm of the Bayes factor in favor of the non-GR hypothesis.}
    \label{fig:nr_wave_sys}
\end{figure*}

\begin{figure*}[!htbp]
    \centering
    \includegraphics[trim={0.4cm 0.4cm 0.4cm 0.3cm},width=0.99\linewidth]{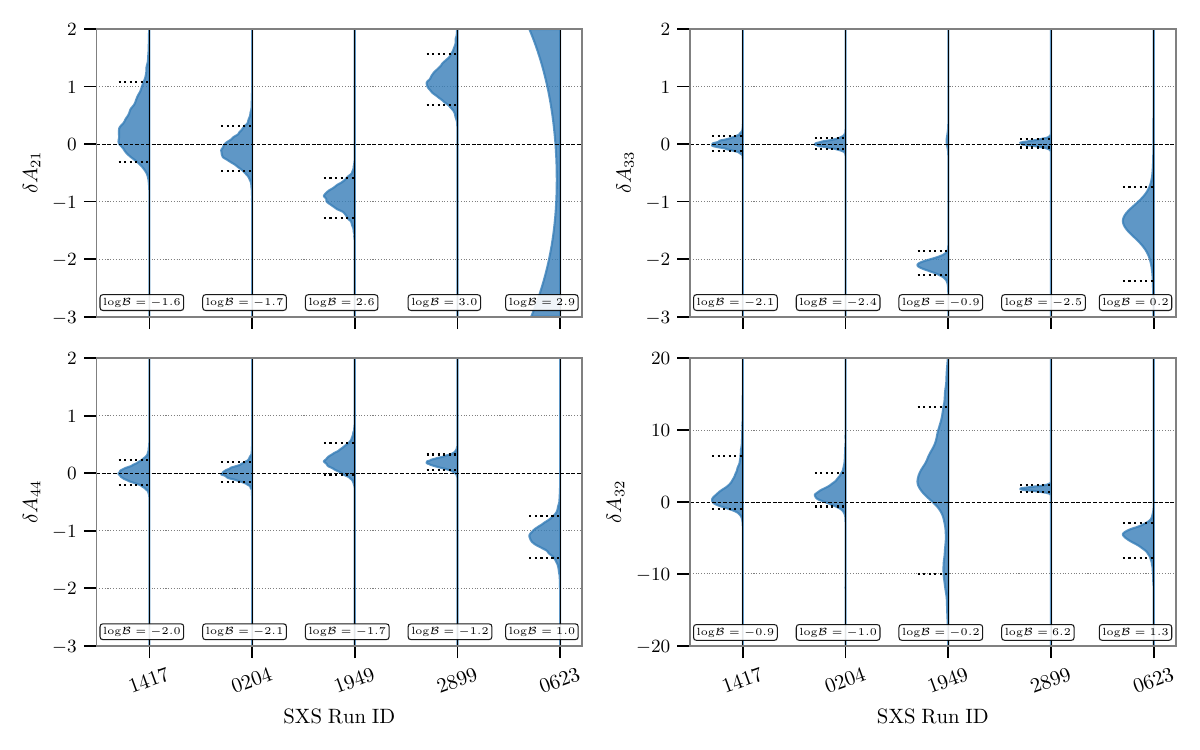}
    \caption{Same as Fig.~\ref{fig:nr_wave_sys} but for $200\,\msun$ systems.}
    \label{fig:nr_wave_sys_hi}
\end{figure*}

\begin{table}[!htbp] 
  \centering
  \caption{\label{tab:nr_runs}Description of the numerical relativity simulations from the SXS catalog. Each system is simulated with $\iota = 60^{\circ}$ and is placed at a distance ensuring an optimal SNR of $\sim 50$ in a LIGO-Hanford, LIGO-Livingston, and Virgo detector network. The lower frequency cutoff is set to 20 Hz to avoid exceeding the simulation’s inspiral cycles.}
    \begin{tabular}{ P{2cm}  P{1cm} P{0.6cm}  P{1cm} P{1cm} P{1cm} P{1.2cm}}
    \hhline{=======}
    SXS ID & $M_{tot}$ & $q$ & $\chi_{\rm eff}$ & $\chi_{\perp,1}$ & $\chi_{\perp,2}$ & $D_L$ \\
    & $(\msun)$ & & & & & (Mpc) \\
    \hhline{-------}
    SXS:BBH:0045 & 100 & 3.0 & 0.25 & 0 & 0 & 689\\
    SXS:BBH:0061 & 100 & 5.0 & 0.42 & 0 & 0 & 570\\
    SXS:BBH:1156 & 100 & 4.4 & 0.33 & 0.27 & 0.76 & 582\\
    SXS:BBH:1410 & 100 & 4.0 & 0.25 & 0.40 & 0.36 & 494\\
    SXS:BBH:1417 & 200 & 4.0 & 0.42 & 0 & 0 & 871\\
    SXS:BBH:0204 & 200 & 7.0 & 0.35 & 0 & 0 & 544\\
    SXS:BBH:1949 & 200 & 4.0 & 0.00 & 0.40 & 0.00 & 613\\
    SXS:BBH:2899 & 200 & 7.0 & 0.55 & 0.43 & 0.53 & 380\\
    SXS:BBH:0623 & 200 & 1.1 & 0.00 & 0.89 & 0.89 & 1310\\
    \hhline{=======}
    \end{tabular}
\end{table}

We perform the SMA test by varying a single $\delta A_{\ell m}$ at a time, using a uniform prior $\delta A_{\ell m}\sim\mathcal{U}[-20,20]$. For $M_{\rm tot}=100\msun$, the posteriors and the corresponding $\log \mathcal{B}^{\rm non\text{-}GR}_{\rm GR}$ values (abbreviated as $\log \mathcal{B}$) are shown in Fig.~\ref{fig:nr_wave_sys}. In all four simulations, the posteriors for every $(\ell,m)$ include $\delta A_{\ell m}=0$ within the 90\% CI, and $\log \mathcal{B}<0$ favors the GR hypothesis. For $M_{\rm tot}=200\msun$, Fig.~\ref{fig:nr_wave_sys_hi} shows significant apparent deviations in multiple cases (see Appendix~\ref{appsec:bias_mitigation} for the recovery of intrinsic parameters). Particularly, pronounced deviations are observed in the $\delta A_{21}$ distribution for SXS:BBH:1949 and SXS:BBH:2899, both with significant mass asymmetry and spin-orbit precession. We also find deviations in $\delta A_{33}$ for SXS:BBH:1949, and $\delta A_{44}$ and $\delta A_{32}$ for SXS:BBH:2899. In fact, for SXS:BBH:2899, the $\delta A_{32}$ yields $\log \mathcal{B} > 5$, which constitutes strong evidence in favor of the non–GR model.

Figure~\ref{fig:nr_wave_sys_hi} also reports results for SXS:BBH:0623. Although $q\approx 1$, the large in–plane spin components produce strong precession, which can enhance HOMs. The source parameters are close to those inferred for GW231123~\cite{LIGOScientific:2025rsn}, where the $(3,3)$ mode was reported together with substantial waveform model-dependent biases in parameter estimates. Consistently, our SXS:BBH:0623 runs indicate that IMRPhenomXPHM modeling errors bias the recovered $\delta A_{\ell m}$, with most posteriors localized, but shifted away from the GR value, and $\delta A_{21}$ railing against the prior boundary $[-20,20]$, with minimal support for $\delta A_{21} = 0$.  

Overall, the SMA test supports the GR hypothesis for most cases, with apparent deviations confined to simulations involving significant in-plane spin and high total mass. These results caution against using SMA with IMRPhenomXPHM for massive, strongly precessing binaries, and motivate implementing SMA with more accurate precessing models in this regime, such as NRSur7dq4~\cite{Varma:2019csw}.

\subsubsection{Effect of orbital eccentricity} \label{subsubsec:ecc}

Orbital eccentricity can significantly alter the structure of gravitational radiation giving rise to additional eccentric harmonics that scale differently from the standard mass-asymmetry–induced modes~\cite{Yunes:2009yz,Tanay:2016zog,Bose:2021pcw}. Since the IMRPhenomXPHM waveform used in the SMA test assumes quasi-circular orbits, it does not model these eccentric contributions. Thus, using this waveform for parameter estimation will give biased inference for parameters (see Appendix~\ref{app:spins}). Consequently, when the SMA test is applied to an eccentric signal, the eccentricity-induced harmonics may be absorbed as amplitude deviations in the standard HOMs, potentially mimicking a violation of GR.

To assess this, we apply the SMA test to two eccentric numerical-relativity simulations from the SXS catalog, SXS:BBH:1371 and SXS:BBH:1373, described in Ref.~\cite{Narayan:2023vhm}. The systems considered are non-spinning binary black holes with a mass ratio of $q = 3$ and a redshifted total mass of $80~M_{\odot}$. Both binaries are observed face-on $(\iota = 0^{\circ})$ at a luminosity distance of $400~\mathrm{Mpc}$. In a three-detector LIGO–Virgo network, these systems yield a network SNR of $\sim87$. The corresponding eccentricities, measured at a GW frequency of $17~\mathrm{Hz}$ for an $80~M_{\odot}$ binary, are $e\simeq0.05$ for SXS:BBH:1371 and $e\simeq0.1$ for SXS:BBH:1373~\cite{Hinder:2017sxy}.

\begin{figure*}[!htbp]
    \centering
    \includegraphics[trim={0.4cm 0.4cm 0.4cm 0.3cm},width=0.8\linewidth]{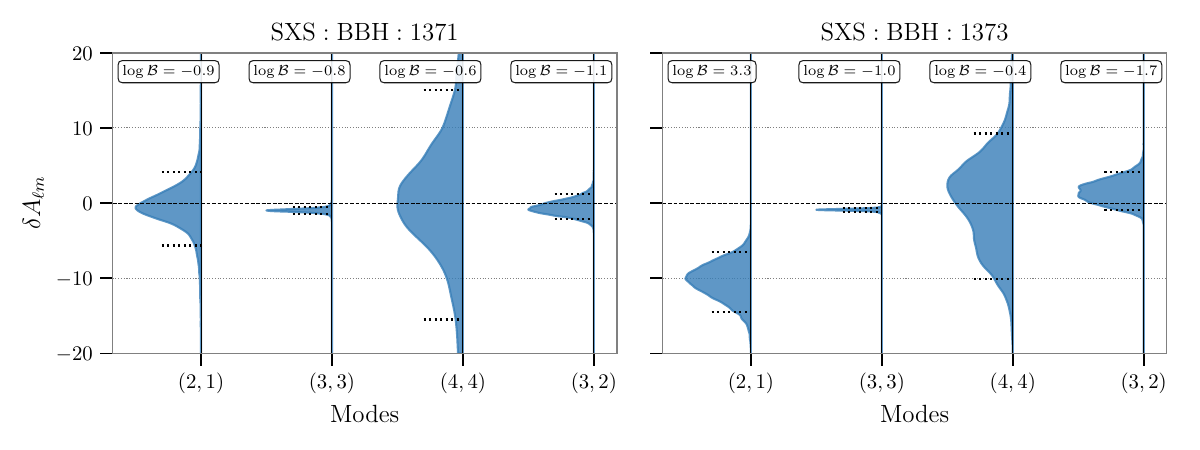}
    \caption{Posterior distributions for $\delta A_{\ell m}$ when parameter estimation is performed on simulated eccentric signals using SXS simulations, SXS:BBH:1371 ($e = 0.055$) and SXS:BBH:1373 ($e = 0.093$). The quoted eccentricities are from Ref.~\cite{Hinder:2017sxy}, evaluated at a PN velocity squared of  $v^2 = 0.075$, corresponding to a dominant $|m| = 2$ GW frequency of $\sim 17~\mathrm{Hz}$ for the redshifted total mass of the $80~M_{\odot}$ binaries considered here. Also mentioned are the corresponding log Bayes factors in favor of the non-GR hypothesis.}
    \label{fig:nr_ecc}
\end{figure*}

By varying one $\delta A_{\ell m}$ parameter at a time, we evaluate the response of the SMA test to the unmodeled eccentricity effects. The results are presented in Fig.~\ref{fig:nr_ecc}. For both binaries considered, the posteriors of $\delta A_{44}$ and $\delta A_{32}$ are consistent with $\delta A_{\ell m} = 0$ within the $90\%$ CI. For the lower-eccentricity binary (SXS:BBH:1371, $e = 0.055$), $\delta A_{21}$ also agrees with $\delta A_{\ell m} = 0$ within the $90\%$ CI, though a deviation is observed in $\delta A_{33}$. In contrast, for the higher-eccentricity binary (SXS:BBH:1373, $e = 0.093$), both $\delta A_{33}$ and $\delta A_{21}$ show deviations, with the latter exhibiting a more pronounced departure and a $\log \mathcal{B} = 3.3$ in favor of the non-GR hypothesis.

These results demonstrate that orbital eccentricity, if not modeled, can mimic amplitude deviations and lead the SMA test to infer apparent violations of GR. Consequently, care must be taken when applying the SMA framework to events exhibiting evidence for non-negligible eccentricity, as eccentric harmonics can be misinterpreted as subdominant-mode anomalies. More broadly, this highlights the need to interpret GR tests in conjunction with astrophysical information about source dynamics, ensuring that any detected deviation is attributed to genuine departures from GR rather than unmodeled physical effects.


\section{Capturing deviations from general relativity} \label{sec:non-gr}

We now assess the ability of the SMA test to capture deviations when applied to amplitude- and phase-deformed signals. Two classes of simulations are considered: first, systems with explicit amplitude deviations in their HOMs, and second, systems with phase deviations introduced at specific PN orders. The former are generated directly using the SMA pipeline, while the latter employ the TIGER (Test Infrastructure for General Relativity) framework~\cite{Agathos:2013upa,Li:2011cg,Roy:2025gzv}. Section~\ref{subsec:res_amp} presents the injection-recovery results for amplitude-rescaled signals, and Sec.~\ref{subsec:phase_dev} explores the response of the SMA test to phase-modified signals.

\subsection{Rescaled amplitudes} \label{subsec:res_amp}

\begin{figure}[!htbp]
    \centering
    \includegraphics[trim={0.4cm 0.4cm 0.4cm 0.3cm},width=0.99\linewidth]{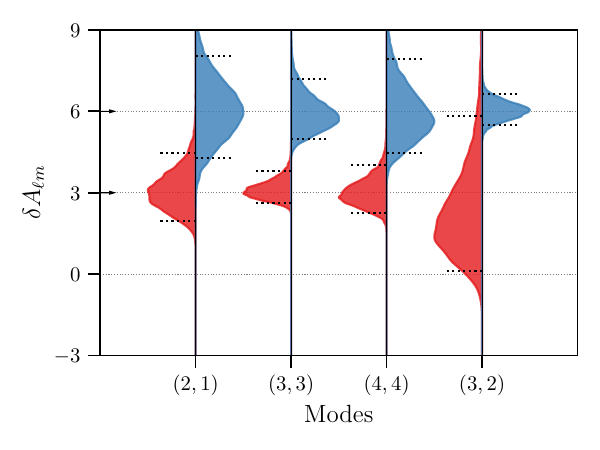}
    \caption{Recovered posteriors for $\delta A_{\ell m}$ from simulated signals with injected deviations $\delta A_{\ell m}=3$ (left, red) and $\delta A_{\ell m}=6$ (right, blue) for each $(\ell,m)$ mode. The dotted horizontal lines indicate the $90\%$ credible intervals. The recovered values agree well with the injections across all modes.}
    \label{fig:non-gr-sma}
\end{figure}

We first evaluate whether the SMA test accurately recovers injected amplitude deviations. To this end, we simulate BBH mergers with a source-frame total mass of $60\msun$, mass ratio $q=4$, and component dimensionless spins $(\chi_1, \chi_2)=(0.2, 0.1)$ aligned with the orbital angular momentum. The signals are injected in zero noise for the LIGO–Hanford, LIGO–Livingston, and Virgo detectors. For each run, we modify the amplitude of one HOM by $\delta A_{\ell m}=3$ or $\delta A_{\ell m}=6$ and recover these deviations using a uniform prior $\mathcal{U}[-10,10]$ on the corresponding $\delta A_{\ell m}$. The luminosity distance $D_L$ is set to 400 Mpc for the $(2,1)$, $(3,3)$, and $(4,4)$ analyses, while for the subdominant $(3,2)$ mode it is reduced to 200 Mpc to ensure adequate SNR in that component. The results are summarized in Fig.~\ref{fig:non-gr-sma}.

\begin{figure*}[!htbp]
    \centering
    \includegraphics[trim={0.4cm 0.4cm 0.4cm 0.3cm},width=0.80\linewidth]{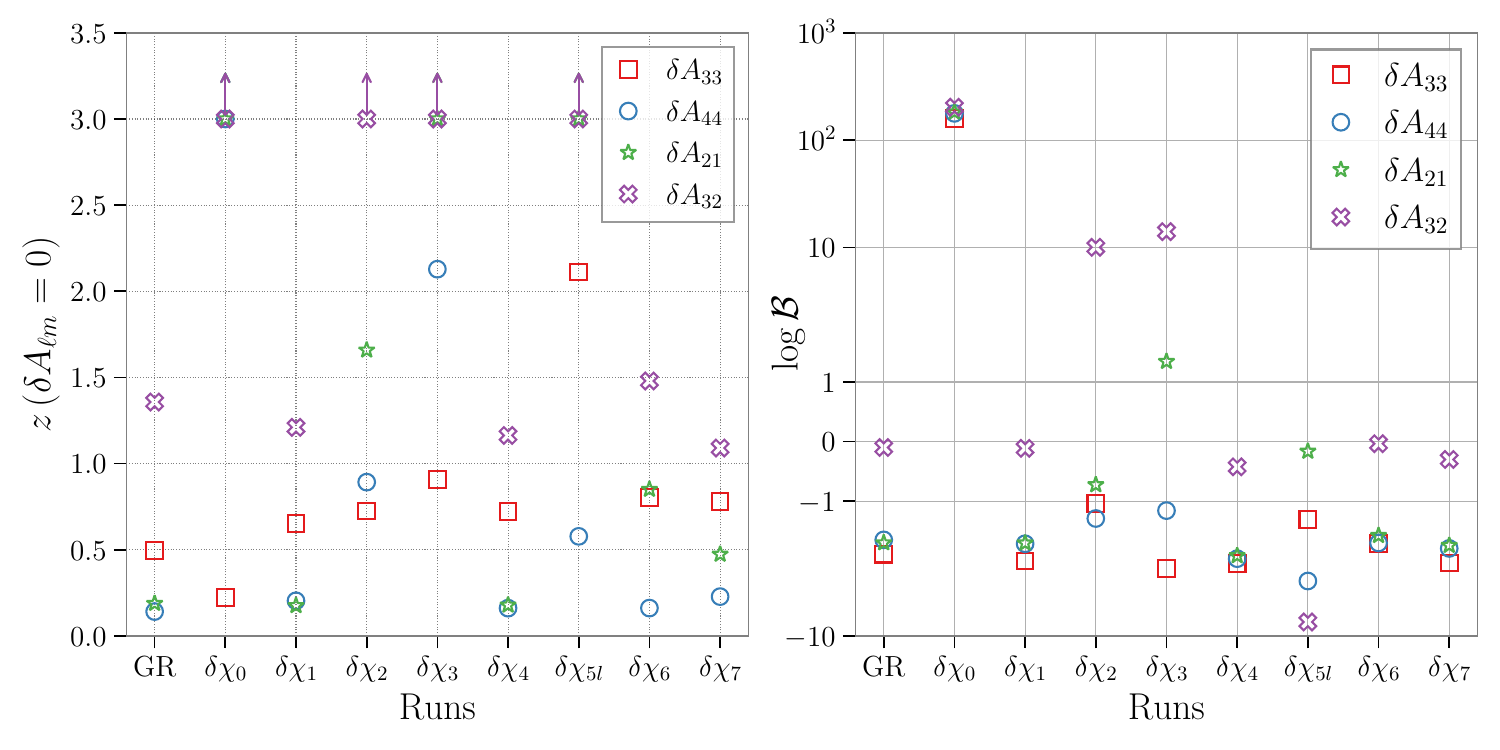}
    \caption{Results of applying the SMA test (one mode at a time) to simulated signals with phase deviations at different PN orders. The left panel shows the $z$-value corresponding to the GR hypothesis ($\delta A_{\ell m}=0$), while the right panel shows the corresponding $\log \mathcal{B}$ in favor of the non-GR hypothesis.}
    \label{fig:non-gr-tiger}
\end{figure*}

The recovered posteriors in Fig.~\ref{fig:non-gr-sma} demonstrate excellent agreement with the injected $\delta A_{\ell m}$ values. For all but one case, the non-GR hypothesis is decisively favored, with $\log \mathcal{B} \gg 10$, corresponding to overwhelming evidence for a deviation from GR. The only exception is the $\delta A_{32}=3$ run, for which $\log \mathcal{B}=-0.5$, indicating no significant preference for the non-GR model. This can be attributed to the broad posterior on $\delta A_{32}$, which retains finite support near $\delta A_{32}=0$ despite excluding it at the $90\%$ level. Increasing the injected deviation to $\delta A_{32}=6$ yields $\log \mathcal{B}=40$, with the posterior sharply centered on the injected value.

To further test mode cross-sensitivity, we also examine whether a deviation, $\delta A_{\ell m}$, injected in one mode can be recovered as a deviation in another when only that mode’s $\delta A_{\ell' m'}$ is allowed to vary. For the $\delta A_{33}=6$ injection, if $\delta A_{21}$ and $\delta A_{44}$ are individually varied, they are recovered as $\delta A_{21}=-0.85 \pm 0.16$ and $\delta A_{44}=-1.04 \pm 0.11$, with corresponding Bayes factors $\log \mathcal{B}=13.6$ and $\log \mathcal{B}=49.2$ in favor of the non-GR hypothesis. This is because all $(\ell,m)$ modes are governed by the same underlying binary parameters; a deviation in one mode can therefore bias the inferred source parameters so as to mimic a deviation in another. Thus, while it perfectly captures deviations it is modeled for, the SMA test can also respond to amplitude deviations not directly associated with the perturbed mode.

\subsection{Phase deviations} \label{subsec:phase_dev}
In the TIGER framework, phase modifications to the inspiral waveform are introduced by perturbing the PN phasing coefficients. At the $(i/2)^{\mathrm{th}}$ PN order, this is expressed as
\begin{equation}
    \phi_i^{\mathrm{GR}}(f) \longrightarrow (1 + \delta \chi_i)\,\phi_i^{\mathrm{GR}}(f),
\end{equation}
where $\delta \chi_i$ quantifies the fractional deviation from the GR value of the $i^{\mathrm{th}}$ phasing term.

Following the setup in Sec.~\ref{subsec:res_amp}, we simulate BBH mergers with the same intrinsic parameters: $M_{\mathrm{tot}}=60\msun$, $q=4$, and component spins $(\chi_1, \chi_2)=(0.2, 0.1)$ aligned with the orbital angular momentum, placed in zero noise for the LIGO–Hanford, LIGO–Livingston, and Virgo detectors. The sources are fixed at $D_L \sim 400$ Mpc, corresponding to a network SNR of 50 for the GR signal. We then generate signals with $\delta \chi_i = 1$ for $i \in \{0, 1, 2, 3, 4, 5l, 6, 7\}$\footnote{Here, $\delta\chi_{5l}$ refers to the logarithmic term at 2.5PN order in the TIGER parametrization.}, and perform the SMA test on each by allowing one $\delta A_{\ell m}$ parameter at a time to vary within a uniform prior $\mathcal{U}[-20,20]$. The results are summarized in Fig.~\ref{fig:non-gr-tiger}.

The left panel of Fig.~\ref{fig:non-gr-tiger} shows the $z$-value of $\delta A_{\ell m}=0$ within the recovered posteriors, while the right panel displays the corresponding $\log \mathcal{B}$ in favor of the non-GR model. The SMA test exhibits strong sensitivity to large phase deviations, particularly at the leading PN order. For $\delta \chi_0=1$, all HOMs yield $\log \mathcal{B} > 100$, decisively favoring the non-GR hypothesis. In these cases, the recovered $\delta A_{\ell m}$ posteriors exclude the GR value beyond $3\sigma$ for all modes except $(3,3)$, whose distribution shows bimodality and significant density near the prior boundaries, producing a comparatively low $z$-value. 

Among all modes, $\delta A_{32}$ emerges as the most responsive to phase deviations. It yields $\log \mathcal{B} > 10$ for both the $\delta \chi_2$ and $\delta \chi_3$ injections, and excludes $\delta A_{32}=0$ beyond $3\sigma$. The $\delta A_{21}$ and $\delta A_{32}$ parameters also show apparent deviations for the $\delta \chi_{5l}$ case, though the associated Bayes factors do not favor the non-GR model. Importantly, for the GR-consistent injections ($\delta \chi_i=0$, denoted by `GR' in the figure), all $\delta A_{\ell m}$ distributions contain zero within $2\sigma$, with $\log \mathcal{B}\lesssim0$, confirming the test’s robustness against false positives.

As with amplitude deviations, phase perturbations bias the inferred binary parameters, which in turn can manifest as apparent amplitude deviations captured by the SMA test. These results demonstrate that, although the SMA framework explicitly models only amplitude modifications, it is also inherently sensitive to phase deviations through their induced parameter correlations. Notably, the $\delta A_{32}$ parameter stands out as particularly effective in detecting such indirect signatures of non-GR phasing.


\section{Applying the SMA test to GW events} \label{sec:gw_events}
We now apply the SMA test to selected GW events from O4. This section presents constraints on the deviation parameters $\delta A_{\ell m}$ for HOM that contribute significantly to the observed signal (see appendix~\ref{app:selection}). In Sec.~\ref{subsec:the_good}, we report results for $\delta A_{33}$ and $\delta A_{44}$ using GW241011~\cite{LIGOScientific:2025brd} and GW230814~\cite{2025arXiv250907348T}, respectively. Section~\ref{subsec:the_bad} examines $\delta A_{44}$ constraints for GW250114~\cite{LIGOScientific:2025rid,
LIGOScientific:2025obp}, where, despite a favorable $\rho^{\perp}_{44}$ distribution, parameter degeneracies, Gaussian noise effects, and possibly waveform systematics, render the posterior largely uninformative. Finally, Sec.~\ref{subsec:the_ugly} presents results for GW231123~\cite{LIGOScientific:2025rsn}, showing that systematic errors in waveform modeling can mimic significant deviations from GR.

\subsection{The good: GW230814 and GW241011} \label{subsec:the_good}

During O4, the LVK detectors observed GW230814 and GW241011, both with network SNRs $\gtrsim 40$. The strong mass asymmetry of GW241011, combined with its high SNR, leads to a prominent contribution from the $(3,3)$ mode, with an orthogonal SNR of $\rho^{\perp}_{33}=5.11^{+1.70}_{-1.52}$. In contrast, GW230814 is a more symmetric system observed close to edge-on, resulting in a measurable $(4,4)$-mode contribution with $\rho^{\perp}_{44}=3.39^{+0.41}_{-1.26}$. These properties make the two events ideal for constraining amplitude deviations in the $(3,3)$ and $(4,4)$ modes, respectively.

\begin{figure}[!htbp]
    \centering
    \includegraphics[trim={0.4cm 0.4cm 0.4cm 0.3cm},width=0.7\linewidth]{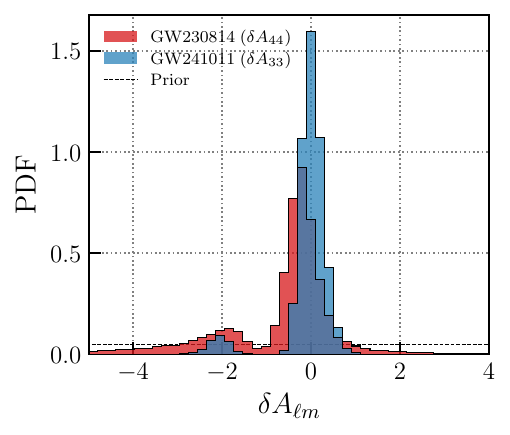}
    \caption{Posterior distributions of $\delta A_{33}$ and $\delta A_{44}$ obtained from GW241011 and GW230814, respectively. The dashed black lines denote the uniform priors $\mathcal{U}[-10,10]$ adopted in each analysis. The dominant mode in both cases is consistent with GR $(\delta A{\ell m} = 0)$, and the second mode at $-2$ can be attributed to parameter degeneracy (see Appendix~\ref{app:degeneracy}).}
    \label{fig:the-good}
\end{figure}

Figure~\ref{fig:the-good} summarizes the resulting posteriors on $\delta A_{33}$ and $\delta A_{44}$. For GW241011, we find $\delta A_{33}=0.00^{+0.46}_{-1.82}$,\footnote{This analysis was first performed by the authors in Ref.~\cite{LIGOScientific:2025brd}, and is repeated here with public data.} with a corresponding $\log\mathcal{B}=-1.3$, indicating preference for the GR hypothesis. The posterior distribution is mildly bimodal, featuring a dominant peak near $\delta A_{33}=0$ and a secondary one near $\delta A_{33}\simeq -2$. This structure arises from the degeneracy between $\delta A_{33}$ and the reference phase, discussed in Sec.~\ref{subsec:gauss_noise} and Appendix~\ref{app:degeneracy}.

For GW230814, we obtain $\delta A_{44}=-0.30^{+1.16}_{-3.45}$, marking one of the first, and the strongest, constraint on the amplitude deviation of the hexadecapolar mode. The posterior exhibits similar features to the $\delta A_{33}$ distribution for GW241011, with a primary peak consistent with GR and a secondary feature near $\delta A_{44}\simeq -2$, again associated with phase degeneracy. The broader width of the posterior reflects the lower SNR in the $(4,4)$ mode. The resulting Bayes factor, $\log\mathcal{B}=-1.1$, likewise favors the GR hypothesis.

Previous analyses, such as Ref.~\cite{Puecher:2022sfm}, employed GW190412 and GW190814 to constrain $\delta A_{33}$, but both exhibited pronounced bimodalities that led to broad and weak bounds on the parameter. In contrast, GW241011 and GW230814 yield the most stringent constraints to date on $\delta A_{33}$ and $\delta A_{44}$, respectively, owing to their higher SNRs and stronger higher-order mode content.

\subsection{The bad: GW250114} \label{subsec:the_bad}

GW250114 is a nearly mass-symmetric BBH merger detected with a network SNR of $\sim80$ and a reported orthogonal SNR of $\rho^{\perp}_{44}=3.6^{+1.4}_{-1.5}$~\cite{KAGRA:2025oiz}. Reanalyzing the event with the IMRPhenomXPHM waveform, we obtain a consistent estimate of $\rho^{\perp}_{44}=3.90^{+1.48}_{-1.57}$. Although this nominally indicates measurable $(4,4)$ mode content, the maximum-likelihood value for the IMRPhenomXPHM analysis corresponds to $\rho^{\perp}_{44}\simeq2$, implying that the contribution of this mode may be marginal. We nevertheless apply the SMA test to constrain amplitude deviations in the $(4,4)$ mode.

\begin{figure}[!htbp]
    \centering
    \includegraphics[trim={0.4cm 0.4cm 0.4cm 0.3cm},width=0.7\linewidth]{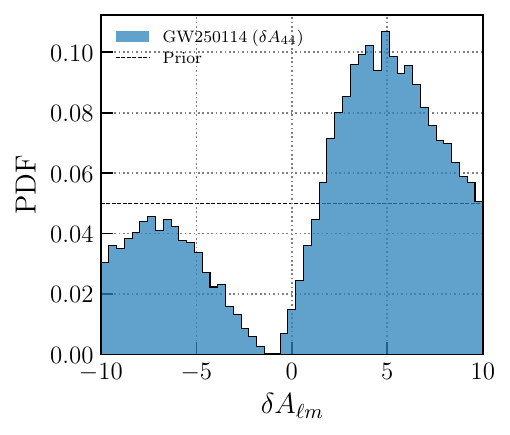}
    \caption{Posterior distribution of $\delta A_{44}$ with GW250114. The dashed black line shows the applied prior, $\mathcal{U}[-10,10]$, for the analyses. We infer an uninformative distribution, with significant support on prior boundaries.}
    \label{fig:the-bad}
\end{figure}

The resulting posterior on $\delta A_{44}$ is shown in Fig.~\ref{fig:the-bad}. The distribution is broad and strongly bimodal, providing little information on the true value of $\delta A_{44}$. This behavior suggests that, at least within the IMRPhenomXPHM model, the $(4,4)$ mode content in GW250114 is insufficient to yield a meaningful constraint on its amplitude deviation. To test this interpretation, we inject a GR-consistent GW250114-like signal, constructed using the maximum-likelihood parameters, into 20 independent Gaussian noise realizations. The recovered posteriors, presented in Fig.~\ref{fig:gauss_c44_250114} of Appendix~\ref{app:degeneracy}, exhibit similarly broad and bimodal structures, confirming that the observed behavior arises from limited mode SNR, gaussian noise effects, and parameter degeneracies, rather than a genuine deviation from GR. Note that a similar analysis performed by ~\citet{Chandra:2025gw250114} using the time-domain waveform model TEOBResumS-Dali~\cite{Gamba:2024cvy,Nagar:2024oyk,Albanesi:2025txj} also recovers a bimodal, but tighter, posterior. While they point to parameter degeneracy, they mainly attribute this result to waveform systematics (however, Ref.~\cite{LIGOScientific:2025rid} finds consistent inference of binary properties with different waveforms).

\subsection{The ugly: GW231123} \label{subsec:the_ugly}

We next apply the SMA test to the intermediate-mass binary black hole merger GW231123. The component masses of this system are inferred to exceed $100\,M_{\odot}$, with dimensionless spin magnitudes $\gtrsim 0.8$ and some evidence of strong spin–orbit precession~\cite{LIGOScientific:2025rsn}. Consequently, the inferred mode SNRs span broad ranges, with $\rho^{\perp}_{21}$, $\rho^{\perp}_{33}$, and $\rho^{\perp}_{44}$ extending up to $\sim12$, $\sim20$, and $\sim15$, respectively (c.f. Fig.~\ref{fig:selection}).

The inferred parameters place GW231123 squarely in a region of parameter space where waveform modeling errors are known to be significant~\cite{Dhani:2024jja}. Indeed, in Sec.~\ref{subsec:NR}, we found that the application of SMA on the NR simulation SXS:BBH:0623, whose parameters are comparable to those of GW231123, exhibited strongly biased $\delta A_{\ell m}$ posteriors at higher SNR ($\sim50$). This motivates caution when interpreting results for such systems, even at the observed network SNR of $\sim22$.

\begin{figure}[!htbp]
    \centering
    \includegraphics[trim={0.4cm 0.4cm 0.4cm 0.3cm},width=0.7\linewidth]{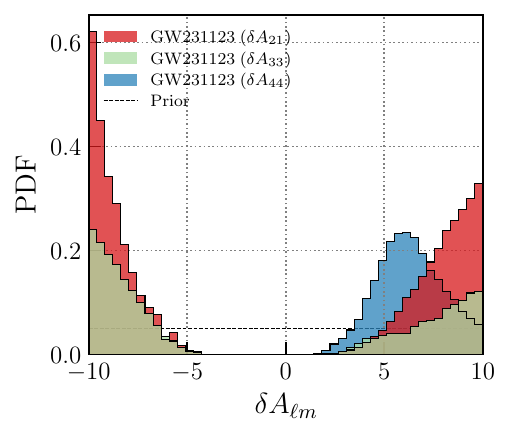}
    \caption{Posterior distributions of $\delta A_{21}$, $\delta A_{33}$, and $\delta A_{44}$ obtained from GW231123. The dashed black line denotes the uniform prior $\mathcal{U}[-10,10]$ used in all three analyses. We obtain no support for the GR hypothesis for all three cases. As this system has a high total mass and exhibits significant spin-orbit precession (similar to the SXS:BBH:0623 case in Sec.~\ref{subsec:NR}), we attribute this result to waveform systematics.}
    \label{fig:the-ugly}
\end{figure}

Figure~\ref{fig:the-ugly} shows the recovered posteriors for $\delta A_{21}$, $\delta A_{33}$, and $\delta A_{44}$. All three distributions are broad, with no support for the GR value $\delta A_{\ell m}=0$. The $\delta A_{21}$ and $\delta A_{33}$ posteriors pile up against the prior boundaries, suggesting that even larger deviations ($|\delta A_{\ell m}|>10$) would be preferred, if allowed. 

GW231123 exemplifies the emerging regime in which modeling uncertainties exceed statistical errors, allowing systematic biases to masquerade as departures from GR. As emphasized in Ref.~\cite{LIGOScientific:2025rsn}, among current waveform models, only NRSur7dq4~\cite{Varma:2019csw} achieves sufficient fidelity for such systems, though even it is not calibrated for component spin magnitudes above $0.8$. This underscores the need for caution when applying precision GR tests to high-mass, strongly precessing binaries, where waveform inaccuracies can mimic genuine deviations from GR.


\section{Conclusions} \label{sec:concl}

We have developed and applied an improved subdominant mode ampltiude test of GR, designed to probe amplitude-level deviations in the higher-order modes of GW signals from binary black hole mergers while keeping the dominant quadrupole mode fixed. In contrast to previous implementations, which allowed deviations only in the $(2,1)$ and $(3,3)$ modes and were therefore limited to systems with significant mass asymmetry or inclined orientations, the present work extends the formalism to include the $(3,2)$ and $(4,4)$ modes. This extension substantially broadens the applicability of the test to more mass-symmetric and face-on binaries, enabling amplitude-based tests of GR across a larger fraction of the observed binary black hole population. Through extensive injection studies, numerical-relativity analyses, and applications to real GW detections, we have established both the statistical reliability and the physical interpretability of the test.

By simulating signals in Gaussian-noise realizations, we verified that the SMA test recovers unbiased posteriors for GR-consistent signals. In these runs, we observed bimodalities in the posterior distributions of the $(3,3)$ deviation parameters, arising from their degeneracy with the reference orbital phase and inclination. This degeneracy disappears when the $(3,3)$-mode SNR is sufficiently high, leading to a unimodal reference-phase distribution and consequently unimodal $\delta A_{\ell m}$ posteriors for the other HOMs. When applied to numerical-relativity simulations, the SMA test remains robust for aligned-spin and mildly precessing systems but exhibits measurable biases when strong spin–orbit precession or orbital eccentricity are present. Both effects can induce apparent deviations in $\delta A_{\ell m}$ even for GR-consistent signals, a behavior that becomes more pronounced for high-mass systems. These findings are consistent with recent studies showing that waveform modeling errors can produce substantial parameter biases in this region of the binary parameter space, emphasizing the need for improved waveform accuracy to ensure reliable tests of gravity.

We also demonstrated that, although the SMA framework explicitly perturbs mode amplitudes, it is inherently sensitive to phase deviations through correlations between binary parameters and mode amplitudes. When applied to phase-modified waveforms generated with the TIGER framework, the SMA test not only recovered nonzero estimates of $\delta A_{\ell m}$ but also yielded strong Bayes factors in favor of the non-GR hypothesis, particularly for deviations introduced at low post-Newtonian orders. Among all modes, the $(3,2)$ mode exhibited the highest sensitivity to such perturbations. This dual response to both amplitude and phase modifications highlights the broader diagnostic power of the SMA framework as a unified tool for detecting diverse manifestations of deviations from GR.

Applying the SMA test to a subset of LVK detections in the fourth observing run, we obtained empirical bounds on higher-order mode amplitudes across a range of binary configurations. The events GW241011 and GW230814, both with network SNRs around 40, provided ideal test cases owing to their significant higher-order mode content. For GW241011, whose strong mass asymmetry enhances the $(3,3)$ mode contribution, we constrained the amplitude deviation to $\delta A_{33}=0.00^{+0.46}_{-1.82}$, finding no evidence for departures from GR. For GW230814, a more symmetric, nearly edge-on system, we measured $\delta A_{44}=-0.30^{+1.16}_{-3.45}$, marking the strongest constraint on the amplitude deviation of the hexadecapolar $(4,4)$ mode till date. Both results are consistent with GR and represent the tightest bounds to date on the respective subdominant-mode amplitudes.  

In contrast, the high-SNR but nearly symmetric mass event GW250114 produced an uninformative, bimodal posterior for $\delta A_{44}$ despite a nominally strong $(4,4)$-mode SNR. This behavior arises from Gaussian noise effects and parameter degeneracies that dominate when the true $(4,4)$ content is marginal, as confirmed through Gaussian-noise injection studies. Finally, the analysis of the intermediate-mass binary GW231123 revealed pronounced apparent deviations across multiple modes: $\delta A_{21}$, $\delta A_{33}$, and $\delta A_{44}$, driven by waveform systematics in a regime of large total mass and strong spin–orbit precession. These findings emphasize that even with current-generation detectors, systematic uncertainties can exceed statistical errors, underscoring the necessity of high-fidelity precessing models to avoid misinterpreting modeling artifacts as genuine violations of GR.

Taken together, these results establish the SMA test as a robust and empirically validated
consistency test of GR. Beyond extending the original SMA formalism, this work introduced a systematic framework for evaluating the reliability of GW-based tests of GR by disentangling the impacts of physical effects, statistical fluctuations, and waveform modeling systematics. Through controlled injections, numerical-relativity comparisons, and real-event analyses, we showed that SMA can remain both sensitive and interpretable across a wide range of binary configurations. Looking ahead, extending the SMA implementation to additional waveform families and performing hierarchical analyses across multiple detections may further enhance its precision and reach, enabling more stringent and comprehensive tests of GR with upcoming detector upgrades and next-generation GW observatories.


\begin{acknowledgments}
The authors would like to thank NV Krishnendu for useful comments on the draft. IG is grateful to Thomas Callister for sharing his plotting scripts which have been used to generate some of the prettier plots in this work. IG would also like to thank Thomas Callister, Rossella Gamba, and Koustav Chandra for discussions regarding the effect of parameter degeneracies. 
IG acknowledges support from the Network for Neutrinos,
Nuclear Astrophysics, and Symmetries (N3AS) Collaboration, NSF grant: PHY-2020275. IG and BS would also like to acknowledge the support of the National Science Foundation via NSF grants: PHY-2207638, AST-2307147, PHY-2308886, PHY-2309064. PN would like to acknowledge the support from the NSF grant: PHY-2308887. LL would like to acknowledge support from UK Royal Society grant ({URF{\textbackslash}R1{\textbackslash}211451}). The authors would also like to acknowledge the LIGO Laboratory computing resources supported by NSF grants:PHY-0757058 and PHY-0823459, and the Gwave cluster maintained by the Institute for Computational and Data Sciences at Penn State University, supported by NSF grants: OAC-2346596, OAC-2201445, OAC- 2103662, OAC-2018299, and PHY-2110594. ST is supported by the Swiss National Science Foundation Ambizione Grant Number: PZ00P2-202204. This material is based upon work supported by NSF's LIGO Laboratory which is a major facility fully funded by the National Science Foundation.

This work has extensively utilized the following tools and software packages: \texttt{numpy}~\cite{harris2020array}, \texttt{scipy}~\cite{2020SciPy-NMeth}, \texttt{matplotlib}~\cite{Hunter:2007}, \texttt{bilby}~\cite{Ashton:2018jfp}, \texttt{bilby pipe}~\cite{bilby_pipe_paper}, \texttt{bilby tgr}~\cite{2024zndo..10940210A}, \texttt{pesummary}~\cite{Hoy:2020vys}, \texttt{dynesty}~\cite{Speagle:2019ivv,sergey_koposov_2025_17268284}, and \texttt{ColorBrewer}~\cite{Harrower01062003}.

\end{acknowledgments}


\appendix

\section{Selection of events} \label{app:selection}
The SMA test will result in informative posterior distributions on $\delta A_{\ell m}$ only when there is significant SNR in the $(\ell,m)$ mode. If the $(\ell,m)$ mode is not present in data, $\rho^{\perp}_{\ell m}$, in Gaussian noise, follows a $\chi$ distribution with two degrees of freedom~\citep{Fairhurst:2019vut,Mills:2020thr,LIGOScientific:2020stg}. A significant distribution from this distribution would indicate the presence of the mode. 

\begin{figure}[!htbp]
    \centering
    \includegraphics[trim={0.4cm 0.4cm 0.4cm 0.3cm},width=0.9\linewidth]{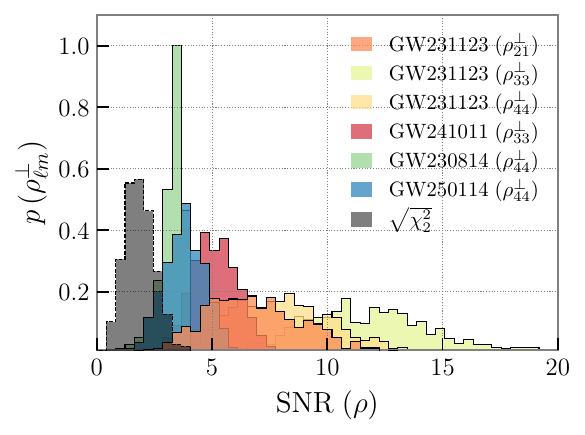}
    \caption{The probability distribution for $\rho^{\perp}_{\ell m}$ for a subset of events detected during O4a. The grey curve shows the $\chi$ distribution with 2 degrees of freedom, representing the null distribution for $\rho^{\perp}_{\ell m}$ if the $(\ell, m)$ mode is not present in data.}
    \label{fig:selection}
\end{figure}

When applied as part of the suite of tests of GR presented by the LVK collaboration~\cite{LIGOScientific:2026qni,LIGOScientific:2026fcf,LIGOScientific:2026wpt}, events are selected for this test when the lower bound of the $68\%$ credible interval of the $\rho^{\perp}_{\ell m}$ distribution exceeds $2.145$ (the $90^{\rm th}$ percentile of a $\chi$ distribution)~\cite{LIGOScientific:2026qni}. For reference, the $\rho^{\perp}_{\ell m}$ for a subset of GW events discussed in Sec.~\ref{sec:gw_events} is shown in Fig.~\ref{fig:selection}, along with the $\chi$ distribution with two degrees of freedom.

\section{Bimodality in the $\delta A_{\ell m}$ posteriors} \label{app:degeneracy}

\begin{figure*}[!htbp]
    \centering
    \includegraphics[trim={0.4cm 0.4cm 0.4cm 0.3cm},width=0.8\linewidth]{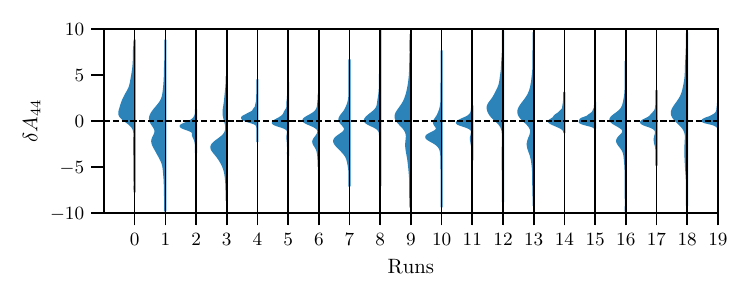}
    \caption{Posterior distributions for $\delta A_{44}$ for GW250114-like systems simulated in 20 Gaussian noise realizations. Several runs show bimodalities, similar to thode observed in Fig.~\ref{fig:gauss_c33_violin}, and when applying the SMA test to GW250114 (c.f. Fig.~\ref{fig:the-bad}).}
    \label{fig:gauss_c44_250114}
\end{figure*}

\begin{figure}[!htbp]
    \centering
    \includegraphics[trim={0.4cm 0.4cm 0.4cm 0.3cm},width=0.95\linewidth]{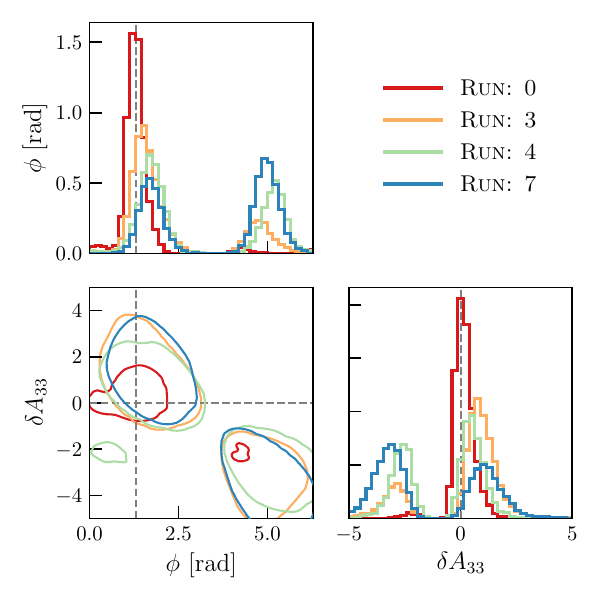}
    \caption{The posterior distributions for the reference phase $\phi_0$ and $\delta A_{33}$ for a subset of Gaussian noise runs. The dashed lines show the injected values for $\phi_0$ and $\delta A_{33}$. We see a strong correlation between the two parameters, with probability density concentrated around $(\delta A_{33},\phi_0) = (1, 1.3)$ and $(-2, 1.3 + \pi)$ regions, where $1.3$ is the injected value of $\phi_0$.}
    \label{fig:degen}
\end{figure}

The bimodalities in the $\delta A_{\ell m}$ posteriors arise from its degeneracy with the inclination angle~\cite{Puecher:2022sfm} and reference orbital phase, and can be exacerbated by Gaussian noise (c.f. Sec.~\ref{subsec:gauss_noise}) and waveform systematics (c.f. Sec.~\ref{subsec:the_ugly}). Specifically, the GW waveform can be written as,
\begin{equation}
    h(t) = \sum_{\ell = 2}^{\infty} \sum_{m=-\ell}^{m=\ell} \mathcal{A}_{\ell m}(t)\,e^{im(\phi(t) + \phi_0)}
\end{equation}
where $\phi_0$ is the reference phase. When the signal only has the $(2,2)$ mode, with negligible contribution from HOMs, the phase term $e^{2i(\phi+\phi_0)}$ has the same value for $\phi$ and $\phi+n\pi$, where $n$ is an odd integer. This degeneracy breaks when the $(3,3)$ mode content in the signal is significant, as for the $(3,3)$ mode, $\phi_0 \rightarrow \phi_0 + n\pi \implies e^{3i(\phi+\phi_0)} \rightarrow -e^{3i(\phi+\phi_0)}$, i.e., the amplitude incurs an extra negative sign. Thus, significant $(3,3)$ mode content can reduce the bimodality in the reference orbital phase.

When the SMA test is applied to a system with bimodal distribution for $\phi_0$, and $\delta A_{33}$ is allowed to vary,  $\delta A_{33} = -2$ compensates for the $-$ sign incurred due to $\phi_0 \rightarrow \phi_0 + n\pi$. Thus, $(\delta A_{\ell m} = 0, \phi_0)$ and $(\delta A_{\ell m} = -2, \phi_0 + n\pi)$ are equivalent. This gives rise to the second mode in $\delta A_{33}$ around $-2$, as seen for some cases in the Gaussian noise runs in Fig.~\ref{fig:gauss_c33_violin}. To further validate our reasoning, we choose a subset of results presented in Fig.~\ref{fig:gauss_c33_violin}, namely runs 0, 3, 4 and 7, with varying levels of bimodality in $\delta A_{33}$, and plot the 2-dimensional distributions between $\delta A_{33}$ and $\phi_0$ in Fig.~\ref{fig:degen}. We note a strong correlation between $\delta A_{33}$ and $\phi_0$, with two distinct concentrations of probability density: around $(\delta A_{33} = 0, \phi_0 = 1.3)$ and $(\delta A_{33} = -2, \phi_0 = 1.3 + \pi)$, where $1.3$ is the simulated value of $\phi_0$.

We note that these bimodalities were not observed for $\delta A_{44}$ in the Gaussian noise runs presented in Sec.~\ref{subsec:gauss_noise}. This is because, for the chosen $48\msun+12\msun$ binary, $D_L$ were set such that $\rho^{\perp}_{44} \sim 3$, but that also meant that $\rho^{\perp}_{33} \sim 8$. Due to the high $(3,3)$ mode SNR, the reference phase distribution was unimodal, and correspondingly no bimodality was seen in the $\delta A_{44}$ posterior. However, a signal with negligible $(3,3)$ mode content but some $(4,4)$ mode content may exhibit bimodalities in $\delta A_{44}$. 

To test this scenario, and to interpret the broad, bimodal $\delta A_{44}$ posterior reported for GW250114 in Sec.~\ref{subsec:the_bad} (see Fig.~\ref{fig:the-bad}), we injected a GW250114-like signal, constructed from the IMRPhenomXPHM maximum-likelihood parameters, into 20 independent Gaussian noise realizations. For this source, the orthogonal SNR in the $(4,4)$ mode is modest, $\rho^{\perp}_{44} \sim 2$, and the $(3,3)$ content is negligible. The recovered posteriors, shown in Fig.~\ref{fig:gauss_c44_250114}, exhibit strong bimodalities in $\delta A_{44}$ across multiple realizations. This confirms that the broad, multi-peaked structure seen in GW250114 is a natural outcome of parameter degeneracy and effect of Gaussian noise in the weak-HOM regime, rather than evidence for a physical deviation from GR.
\begin{figure*}[!htbp]
    \centering
    \includegraphics[trim={0.4cm 0.4cm 0.4cm 0.3cm},width=0.95\linewidth]{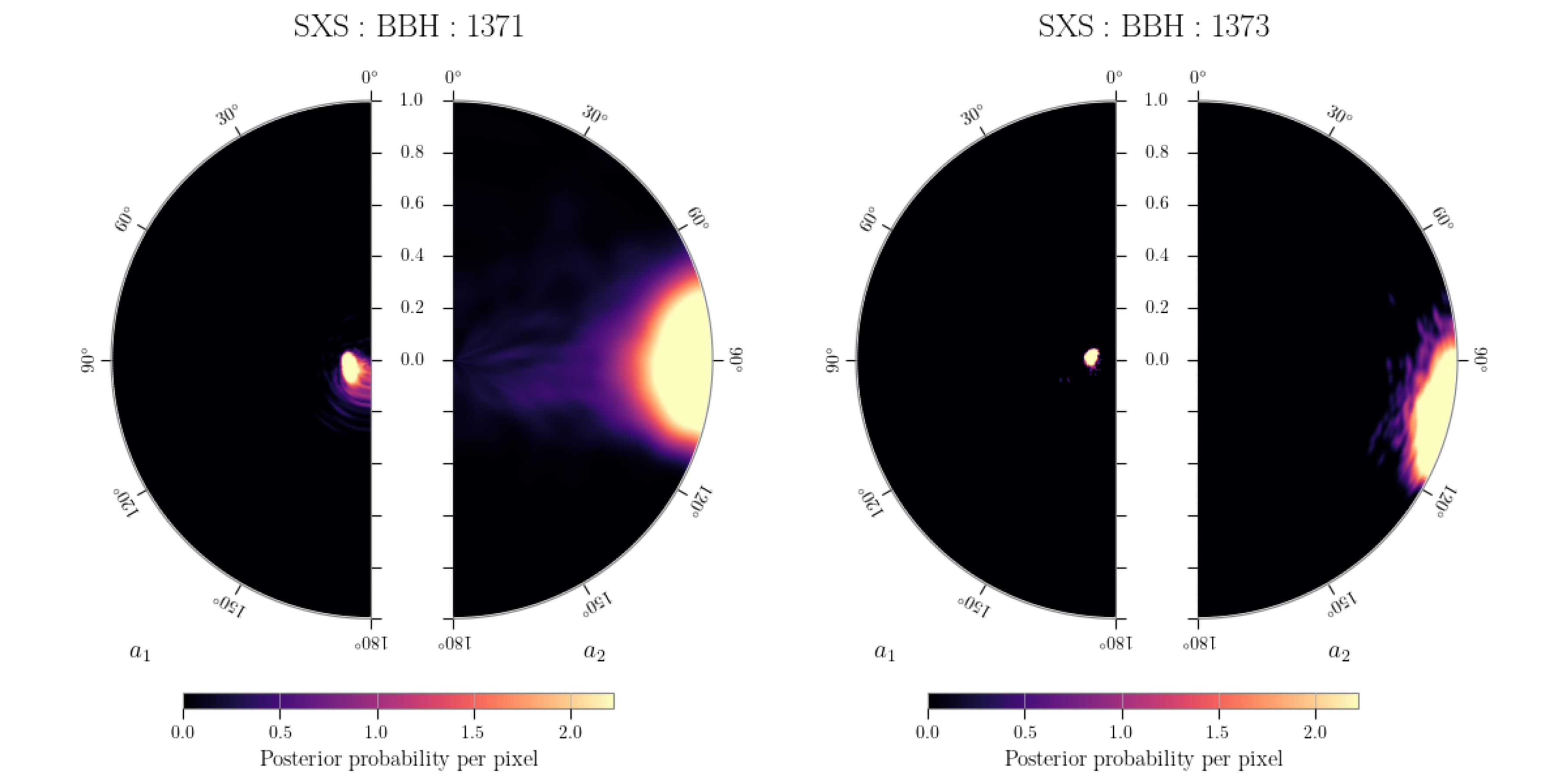}
    \caption{The posterior distributions for primary and secondary spin parameters, when inferred using the GR template applied on SXS:BBH:1371 and SXS:BBH:1373. The radial coordinate denotes the spin magnitude, and the angular component shows the tilt direction.}
    \label{fig:ecc_spin}
\end{figure*}

\section{Extremal spin measurements for eccentric and non-GR runs} \label{app:spins}

When analyzing the eccentric injections (Sec.~\ref{subsubsec:ecc}) and the phase-modified TIGER injections (Sec.~\ref{subsec:phase_dev}), we also performed parameter estimation under the baseline GR hypothesis. In these GR runs, we fixed all SMA deviation parameters to zero (i.e., $\delta A_{\ell m}=0$ for all modes) and recovered only the standard binary parameters using a non-eccentric, precessing waveform model. In this appendix, we focus on an unexpected and repeatable feature of those GR analyses: the secondary black hole is frequently inferred to have a nearly extremal spin.

Figure~\ref{fig:ecc_spin} shows the recovered spin posteriors for the eccentric SXS simulations SXS:BBH:1371 and SXS:BBH:1373, when analyzed in this non-eccentric, GR-consistent manner. Both systems are non-spinning in the numerical-relativity injection, yet the recovered posteriors display the following behavior. First, the primary spin magnitude $a_1$ is inferred to be small, but non-zero. Second, and more surprisingly, the secondary spin magnitude $a_2$ has significant support all the way up to the extremal limit, $a_2 \simeq 1$. In addition, the inferred spin orientations for both components are concentrated in the orbital plane, implying large in-plane spin components and therefore strong spin-orbit precession. Prior studies have shown that orbital eccentricity can be partially mimicked by precession in parameter estimation \cite{Romero-Shaw:2022fbf,Divyajyoti:2025cwq}. However, those works did not explicitly demonstrate posteriors with strong support at (or near) the Kerr extremal limit for the secondary spin. Our results show that this can occur: the model tries to fit to the eccentric features by driving $a_2$ toward unity and tilting both spins into the orbital plane.

We observe a closely related effect in the TIGER injections with modified inspiral phase (Sec.~\ref{subsec:phase_dev}). There, the injected systems have moderate, aligned component spins and GR-consistent amplitudes, but their inspiral phasing is deliberately deformed via nonzero $\delta \chi_i$. When these phase-modified signals are analyzed with the same GR, quasi-circular, precessing template (again with all $\delta A_{\ell m}=0$), the recovered spin of the secondary object again shows substantial support for very large magnitudes, often approaching the extremal limit.

In both the eccentric and phase-deformed cases, near-extremal secondary spin posteriors should therefore not be interpreted as evidence for a rapidly rotating Kerr black hole. Instead, they are a symptom of waveform stress: missing physics (eccentricity in one case, beyond-GR phasing in the other) is being mapped onto the template’s spin degrees of freedom. A full exploration of this effect is beyond the scope of this work. We highlight it here as a cautionary diagnostic for future analyses, and encourage interested readers to explore it further.

\section{Impact of SMA parameters on waveform-systematics-induced parameter biases}
\label{appsec:bias_mitigation}

\begin{figure*}
    \centering
    \includegraphics[width=0.7\textwidth]{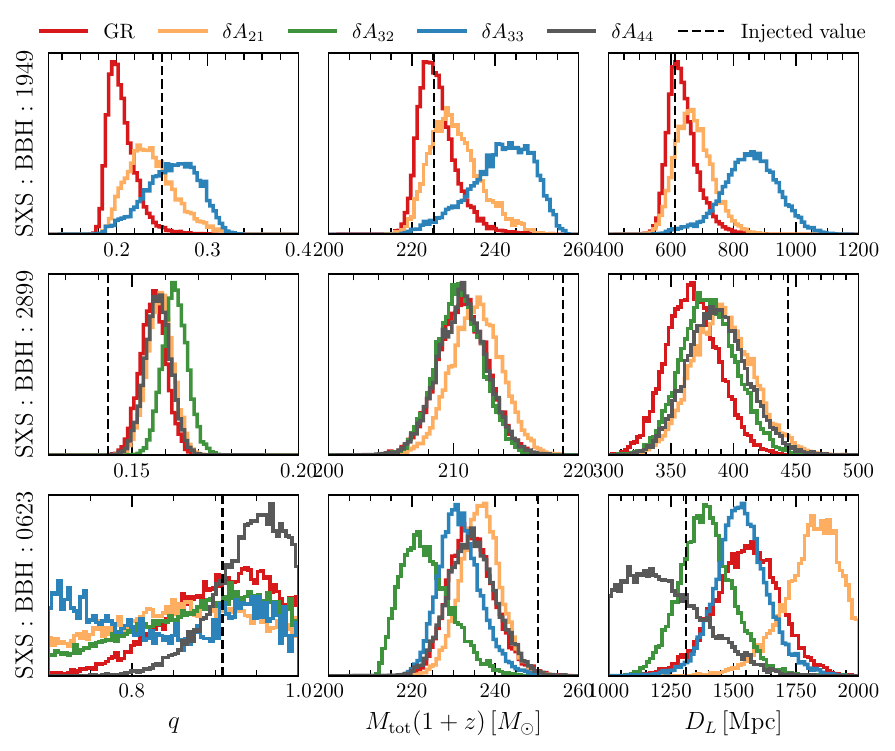}
    \caption{Posterior distributions for the mass ratio $q$, redshifted total mass $M_{\rm tot}(1+z)$, and luminosity distance $D_L$ for three NR injections: SXS:BBH:1949, SXS:BBH:2899, and SXS:BBH:0623 (c.f. Table~\ref{tab:nr_runs} in Sec.~\ref{subsec:NR}). The red curves show the baseline GR recovery, in which all $\delta A_{\ell m}$ are fixed to zero. The remaining curves show SMA recoveries in which one deviation parameter is varied at a time. Only SMA runs for which the corresponding $\delta A_{\ell m}$ posterior showed an apparent deviation from zero in Fig.~\ref{fig:nr_wave_sys_hi} are included. The vertical dashed lines denote the injected values for the respective parameters. Introducing an SMA deviation parameter does not generically mitigate biases in the recovered source parameters. In some cases, such as SXS:BBH:1949, allowing $\delta A_{\ell m}$ to vary can shift the inferred $M_{\rm tot}(1+z)$ and $D_L$ farther from the injected value.
    }
    \label{fig:bias_mitigation}
\end{figure*}

In Sec.~\ref{subsubsec:wave_sys}, we showed that applying the SMA test to NR injections can lead to apparent nonzero values of $\delta A_{\ell m}$ when the recovery waveform does not accurately describe the injected signal. This is apparent for massive, strongly precessing systems, where waveform systematics in IMRPhenomXPHM become appreciable. A natural question is whether the additional SMA degree of freedom may also mitigate the corresponding biases in the recovered intrinsic GR parameters. In other words, if the deviation parameter absorbs part of the mismatch between the injection and the recovery model, one might expect the posterior of parameters such as mass ratio $q$, total mass $M_{\rm tot}$, or luminosity distance $D_L$, to be consistent with the injected value.

We investigate this possibility for three representative NR injections: SXS:BBH:1949, SXS:BBH:2899, and SXS:BBH:0623. These are cases where at least one of the SMA analyses in Sec.~\ref{subsubsec:wave_sys} shows an apparent deviation from the GR value (c.f. Fig.~\ref{fig:nr_wave_sys_hi}). For each injection, we compare the GR recovery, in which all $\delta A_{\ell m}$ are fixed to zero, with SMA recoveries in which one deviation parameter is varied at a time. Figure~\ref{fig:bias_mitigation} shows the marginalized posteriors for $q$, redshifted total mass $M_{\rm tot}(1+z)$, and luminosity distance $D_L$.

The results show that introducing $\delta A_{\ell m}$ does not generically reduce intrinsic parameter biases. For SXS:BBH:1949, allowing a nonzero subdominant-mode amplitude can make posteriors for some parameters (for e.g., $q$) more consistent with the injected value, but it can also introduce additional biases. In particular, the $\delta A_{33}$ recovery shifts the inferred $M_{\rm tot}(1+z)$ and $D_L$ distributions away from the injected values. This demonstrates that the extra degree of freedom can absorb part of the waveform mismatch in a way that improves one aspect of the fit while degrading the recovery of other physical parameters. For SXS:BBH:2899, the recovered posteriors remain biased relative to the injected values across the GR and SMA analyses, with no clear or monotonic improvement when $\delta A_{\ell m}$ is varied. For SXS:BBH:0623, the effect is even less uniform: different choices of $\delta A_{\ell m}$ shift the posteriors in different directions, and no single SMA rescaling consistently improves the recovery of $q$, $M_{\rm tot}(1+z)$, and $D_L$ simultaneously.

This behavior is expected. The mismatch between the injected NR signal and the recovery waveform arises from waveform systematics associated with strong precession and high total mass. Such systematics are not, in general, equivalent to a constant rescaling of a single subdominant-mode amplitude. Whether an additional parameter mitigates a bias depends on how well the new degree of freedom aligns with the missing physical or modeling effect in waveform space.

These results show that allowing $\delta A_{\ell m}$ to vary can change the inferred intrinsic parameters, and, in some cases, may partially reduce biases in individual parameters. However, this behavior is not systematic across events, modes, or parameters. The effect depends on the nature of the mismatch between the injected signal and the recovery model. Thus, we do not interpret the SMA parameters as a general prescription for mitigating biases induced by waveform systematics. Rather, we use them as phenomenological consistency parameters that may reveal when the recovery model does not adequately describe the signal.


\bibliography{bibliography}

@article{Agathos:2013upa,
    author = "Agathos, Michalis and Del Pozzo, Walter and Li, Tjonnie G. F. and Van Den Broeck, Chris and Veitch, John and Vitale, Salvatore",
    title = "{TIGER: A data analysis pipeline for testing the strong-field dynamics of general relativity with gravitational wave signals from coalescing compact binaries}",
    eprint = "1311.0420",
    archivePrefix = "arXiv",
    primaryClass = "gr-qc",
    doi = "10.1103/PhysRevD.89.082001",
    journal = "Phys. Rev. D",
    volume = "89",
    number = "8",
    pages = "082001",
    year = "2014"
}

@article{Ashton:2018jfp,
    author = "Ashton, Gregory and others",
    title = "{BILBY: A user-friendly Bayesian inference library for gravitational-wave astronomy}",
    eprint = "1811.02042",
    archivePrefix = "arXiv",
    primaryClass = "astro-ph.IM",
    doi = "10.3847/1538-4365/ab06fc",
    journal = "Astrophys. J. Suppl.",
    volume = "241",
    number = "2",
    pages = "27",
    year = "2019"
}

@article{Blanchet:2013haa,
    author = "Blanchet, Luc",
    title = "{Post-Newtonian Theory for Gravitational Waves}",
    eprint = "1310.1528",
    archivePrefix = "arXiv",
    primaryClass = "gr-qc",
    doi = "10.12942/lrr-2014-2",
    journal = "Living Rev. Rel.",
    volume = "17",
    pages = "2",
    year = "2014"
}

@article{Boyle:2019kee,
    author = "Boyle, Michael and others",
    title = "{The SXS Collaboration catalog of binary black hole simulations}",
    eprint = "1904.04831",
    archivePrefix = "arXiv",
    primaryClass = "gr-qc",
    doi = "10.1088/1361-6382/ab34e2",
    journal = "Class. Quant. Grav.",
    volume = "36",
    number = "19",
    pages = "195006",
    year = "2019"
}

@article{Chatziioannou:2014bma,
    author = "Chatziioannou, Katerina and Cornish, Neil and Klein, Antoine and Yunes, Nicol\'as",
    title = "{Detection and Parameter Estimation of Gravitational Waves from Compact Binary Inspirals with Analytical Double-Precessing Templates}",
    eprint = "1404.3180",
    archivePrefix = "arXiv",
    primaryClass = "gr-qc",
    doi = "10.1103/PhysRevD.89.104023",
    journal = "Phys. Rev. D",
    volume = "89",
    number = "10",
    pages = "104023",
    year = "2014"
}

@article{Dhani:2024jja,
    author = {Dhani, Arnab and V{\"o}lkel, Sebastian H. and Buonanno, Alessandra and Estelles, Hector and Gair, Jonathan and Pfeiffer, Harald P. and Pompili, Lorenzo and Toubiana, Alexandre},
    title = "{Systematic Biases in Estimating the Properties of Black Holes Due to Inaccurate Gravitational-Wave Models}",
    eprint = "2404.05811",
    archivePrefix = "arXiv",
    primaryClass = "gr-qc",
    doi = "10.1103/5pks-qz6b",
    journal = "Phys. Rev. X",
    volume = "15",
    number = "3",
    pages = "031036",
    year = "2025"
}

@article{Gupta:2024bqn,
    author = "Gupta, Ish",
    title = "{Inferring Small Neutron Star Spins with Neutron Star\textendash{}Black Hole Mergers}",
    eprint = "2402.07075",
    archivePrefix = "arXiv",
    primaryClass = "astro-ph.HE",
    doi = "10.3847/1538-4357/ad49a0",
    journal = "Astrophys. J.",
    volume = "970",
    number = "1",
    pages = "12",
    year = "2024"
}

@unpublished{Gupta:2024gun,
    author = "Gupta, Anuradha and others",
    title = "{Possible Causes of False General Relativity Violations in Gravitational Wave Observations}",
    eprint = "2405.02197",
    archivePrefix = "arXiv",
    primaryClass = "gr-qc",
    doi = "10.21468/SciPostPhysCommRep.5",
    month = "5",
    year = "2024"
}

@article{Hinder:2017sxy,
    author = "Hinder, Ian and Kidder, Lawrence E. and Pfeiffer, Harald P.",
    title = "{Eccentric binary black hole inspiral-merger-ringdown gravitational waveform model from numerical relativity and post-Newtonian theory}",
    eprint = "1709.02007",
    archivePrefix = "arXiv",
    primaryClass = "gr-qc",
    doi = "10.1103/PhysRevD.98.044015",
    journal = "Phys. Rev. D",
    volume = "98",
    number = "4",
    pages = "044015",
    year = "2018"
}

@article{Islam:2019dmk,
    author = "Islam, Tousif and Mehta, Ajit Kumar and Ghosh, Abhirup and Varma, Vijay and Ajith, Parameswaran and Sathyaprakash, B. S.",
    title = "{Testing the no-hair nature of binary black holes using the consistency of multipolar gravitational radiation}",
    eprint = "1910.14259",
    archivePrefix = "arXiv",
    primaryClass = "gr-qc",
    reportNumber = "LIGO-P1900318-v2",
    doi = "10.1103/PhysRevD.101.024032",
    journal = "Phys. Rev. D",
    volume = "101",
    number = "2",
    pages = "024032",
    year = "2020"
}

@article{KAGRA:2021vkt,
    author = "Abbott, R. and others",
    collaboration = "KAGRA, VIRGO, LIGO Scientific",
    title = "{GWTC-3: Compact Binary Coalescences Observed by LIGO and Virgo during the Second Part of the Third Observing Run}",
    eprint = "2111.03606",
    archivePrefix = "arXiv",
    primaryClass = "gr-qc",
    reportNumber = "LIGO-P2000318",
    doi = "10.1103/PhysRevX.13.041039",
    journal = "Phys. Rev. X",
    volume = "13",
    number = "4",
    pages = "041039",
    year = "2023"
}

@article{LIGOScientific:2016aoc,
    author = "Abbott, B. P. and others",
    collaboration = "LIGO Scientific, Virgo",
    title = "{Observation of Gravitational Waves from a Binary Black Hole Merger}",
    eprint = "1602.03837",
    archivePrefix = "arXiv",
    primaryClass = "gr-qc",
    reportNumber = "LIGO-P150914",
    doi = "10.1103/PhysRevLett.116.061102",
    journal = "Phys. Rev. Lett.",
    volume = "116",
    number = "6",
    pages = "061102",
    year = "2016"
}

@article{LIGOScientific:2016lio,
    author = "Abbott, B. P. and others",
    collaboration = "LIGO Scientific, Virgo",
    title = "{Tests of general relativity with GW150914}",
    eprint = "1602.03841",
    archivePrefix = "arXiv",
    primaryClass = "gr-qc",
    reportNumber = "LIGO-P1500213",
    doi = "10.1103/PhysRevLett.116.221101",
    journal = "Phys. Rev. Lett.",
    volume = "116",
    number = "22",
    pages = "221101",
    year = "2016",
    note = "[Erratum: Phys.Rev.Lett. 121, 129902 (2018)]"
}

@article{LIGOScientific:2018dkp,
    author = "Abbott, B. P. and others",
    collaboration = "LIGO Scientific, Virgo",
    title = "{Tests of General Relativity with GW170817}",
    eprint = "1811.00364",
    archivePrefix = "arXiv",
    primaryClass = "gr-qc",
    reportNumber = "LIGO-P1800059",
    doi = "10.1103/PhysRevLett.123.011102",
    journal = "Phys. Rev. Lett.",
    volume = "123",
    number = "1",
    pages = "011102",
    year = "2019"
}

@article{LIGOScientific:2020stg,
    author = "Abbott, R. and others",
    collaboration = "LIGO Scientific, Virgo",
    title = "{GW190412: Observation of a Binary-Black-Hole Coalescence with Asymmetric Masses}",
    eprint = "2004.08342",
    archivePrefix = "arXiv",
    primaryClass = "astro-ph.HE",
    reportNumber = "LIGO-P190412",
    doi = "10.1103/PhysRevD.102.043015",
    journal = "Phys. Rev. D",
    volume = "102",
    number = "4",
    pages = "043015",
    year = "2020"
}

@article{LIGOScientific:2020zkf,
    author = "Abbott, R. and others",
    collaboration = "LIGO Scientific, Virgo",
    title = "{GW190814: Gravitational Waves from the Coalescence of a 23 Solar Mass Black Hole with a 2.6 Solar Mass Compact Object}",
    eprint = "2006.12611",
    archivePrefix = "arXiv",
    primaryClass = "astro-ph.HE",
    reportNumber = "LIGO-P190814",
    doi = "10.3847/2041-8213/ab960f",
    journal = "Astrophys. J. Lett.",
    volume = "896",
    number = "2",
    pages = "L44",
    year = "2020"
}

@article{LIGOScientific:2021sio,
    author = "Abbott, R. and others",
    collaboration = "LIGO Scientific, VIRGO, KAGRA",
    title = "{Tests of General Relativity with GWTC-3}",
    eprint = "2112.06861",
    archivePrefix = "arXiv",
    primaryClass = "gr-qc",
    reportNumber = "LIGO-P2100275",
    doi = "10.1103/PhysRevD.112.084080",
    journal = "Phys. Rev. D",
    volume = "112",
    number = "8",
    pages = "084080",
    year = "2025"
}

@article{Mills:2020thr,
    author = "Mills, Cameron and Fairhurst, Stephen",
    title = "{Measuring gravitational-wave higher-order multipoles}",
    eprint = "2007.04313",
    archivePrefix = "arXiv",
    primaryClass = "gr-qc",
    doi = "10.1103/PhysRevD.103.024042",
    journal = "Phys. Rev. D",
    volume = "103",
    number = "2",
    pages = "024042",
    year = "2021"
}

@article{Narayan:2023vhm,
    author = "Narayan, Purnima and Johnson-McDaniel, Nathan K. and Gupta, Anuradha",
    title = "{Effect of ignoring eccentricity in testing general relativity with gravitational waves}",
    eprint = "2306.04068",
    archivePrefix = "arXiv",
    primaryClass = "gr-qc",
    doi = "10.1103/PhysRevD.108.064003",
    journal = "Phys. Rev. D",
    volume = "108",
    number = "6",
    pages = "064003",
    year = "2023"
}

@article{Pan:2010hz,
    author = "Pan, Yi and Buonanno, Alessandra and Fujita, Ryuichi and Racine, Etienne and Tagoshi, Hideyuki",
    title = "{Post-Newtonian factorized multipolar waveforms for spinning, non-precessing black-hole binaries}",
    eprint = "1006.0431",
    archivePrefix = "arXiv",
    primaryClass = "gr-qc",
    doi = "10.1103/PhysRevD.83.064003",
    journal = "Phys. Rev. D",
    volume = "83",
    pages = "064003",
    year = "2011",
    note = "[Erratum: Phys.Rev.D 87, 109901 (2013)]"
}

@article{Pratten:2020ceb,
    author = "Pratten, Geraint and others",
    title = "{Computationally efficient models for the dominant and subdominant harmonic modes of precessing binary black holes}",
    eprint = "2004.06503",
    archivePrefix = "arXiv",
    primaryClass = "gr-qc",
    doi = "10.1103/PhysRevD.103.104056",
    journal = "Phys. Rev. D",
    volume = "103",
    number = "10",
    pages = "104056",
    year = "2021"
}

@article{Puecher:2022sfm,
    author = "Puecher, Anna and Kalaghatgi, Chinmay and Roy, Soumen and Setyawati, Yoshinta and Gupta, Ish and Sathyaprakash, B. S. and Van Den Broeck, Chris",
    title = "{Testing general relativity using higher-order modes of gravitational waves from binary black holes}",
    eprint = "2205.09062",
    archivePrefix = "arXiv",
    primaryClass = "gr-qc",
    reportNumber = "LIGO DCC: LIGO-P2200140",
    doi = "10.1103/PhysRevD.106.082003",
    journal = "Phys. Rev. D",
    volume = "106",
    number = "8",
    pages = "082003",
    year = "2022"
}

@article{Roy:2019phx,
    author = "Roy, Soumen and Sengupta, Anand S. and Arun, K. G.",
    title = "{Unveiling the spectrum of inspiralling binary black holes}",
    eprint = "1910.04565",
    archivePrefix = "arXiv",
    primaryClass = "gr-qc",
    reportNumber = "LIGO-DCC No. P1900257",
    doi = "10.1103/PhysRevD.103.064012",
    journal = "Phys. Rev. D",
    volume = "103",
    number = "6",
    pages = "064012",
    year = "2021"
}

@article{Schmidt:2014iyl,
    author = "Schmidt, Patricia and Ohme, Frank and Hannam, Mark",
    title = "{Towards models of gravitational waveforms from generic binaries II: Modelling precession effects with a single effective precession parameter}",
    eprint = "1408.1810",
    archivePrefix = "arXiv",
    primaryClass = "gr-qc",
    doi = "10.1103/PhysRevD.91.024043",
    journal = "Phys. Rev. D",
    volume = "91",
    number = "2",
    pages = "024043",
    year = "2015"
}

@article{Speagle:2019ivv,
    author = "Speagle, Joshua S.",
    title = "{dynesty: a dynamic nested sampling package for estimating Bayesian posteriors and evidences}",
    eprint = "1904.02180",
    archivePrefix = "arXiv",
    primaryClass = "astro-ph.IM",
    doi = "10.1093/mnras/staa278",
    journal = "Mon. Not. Roy. Astron. Soc.",
    volume = "493",
    number = "3",
    pages = "3132--3158",
    year = "2020"
}

@ARTICLE{2025arXiv250907348T,
       author = {{The LIGO Scientific Collaboration} and {The Virgo Collaboration} and {The Kagra Collaboration} and {Others}},
        title = "{GW230814: investigation of a loud gravitational-wave signal observed with a single detector}",
      journal = {arXiv e-prints},
         year = 2025,
        month = sep,
          eid = {arXiv:2509.07348},
        pages = {arXiv:2509.07348},
          doi = {10.48550/arXiv.2509.07348},
archivePrefix = {arXiv},
       eprint = {2509.07348},
 primaryClass = {gr-qc},
       adsurl = {https://ui.adsabs.harvard.edu/abs/2025arXiv250907348T}
}

@article{Varma:2019csw,
    author = "Varma, Vijay and Field, Scott E. and Scheel, Mark A. and Blackman, Jonathan and Gerosa, Davide and Stein, Leo C. and Kidder, Lawrence E. and Pfeiffer, Harald P.",
    title = "{Surrogate models for precessing binary black hole simulations with unequal masses}",
    eprint = "1905.09300",
    archivePrefix = "arXiv",
    primaryClass = "gr-qc",
    doi = "10.1103/PhysRevResearch.1.033015",
    journal = "Phys. Rev. Research.",
    volume = "1",
    pages = "033015",
    year = "2019"
}

@article{Li:2011cg,
    author = "Li, T. G. F. and Del Pozzo, W. and Vitale, S. and Van Den Broeck, C. and Agathos, M. and Veitch, J. and Grover, K. and Sidery, T. and Sturani, R. and Vecchio, A.",
    title = "{Towards a generic test of the strong field dynamics of general relativity using compact binary coalescence}",
    eprint = "1110.0530",
    archivePrefix = "arXiv",
    primaryClass = "gr-qc",
    doi = "10.1103/PhysRevD.85.082003",
    journal = "Phys. Rev. D",
    volume = "85",
    pages = "082003",
    year = "2012"
}

@unpublished{Roy:2025gzv,
    author = "Roy, Soumen and Haney, Maria and Pratten, Geraint and Pang, Peter T. H. and Van Den Broeck, Chris",
    title = "{An improved parametrized test of general relativity using the IMRPhenomX waveform family: Including higher harmonics and precession}",
    eprint = "2504.21147",
    archivePrefix = "arXiv",
    primaryClass = "gr-qc",
    reportNumber = "LIGO DCC P2500034",
    month = "4",
    year = "2025"
}

@unpublished{LIGOScientific:2025slb,
    author = "Abac, A. G. and others",
    collaboration = "LIGO Scientific, VIRGO, KAGRA",
    title = "{GWTC-4.0: Updating the Gravitational-Wave Transient Catalog with Observations from the First Part of the Fourth LIGO-Virgo-KAGRA Observing Run}",
    eprint = "2508.18082",
    archivePrefix = "arXiv",
    primaryClass = "gr-qc",
    reportNumber = "LIGO-P2400386",
    month = "8",
    year = "2025"
}

@article{Colleoni:2024knd,
    author = "Colleoni, Marta and Vidal, Felip A. Ramis and Garc{\'\i}a-Quir{\'o}s, Cecilio and Ak{\c{c}}ay, Sarp and Bera, Sayantani",
    title = "{Fast frequency-domain gravitational waveforms for precessing binaries with a new twist}",
    eprint = "2412.16721",
    archivePrefix = "arXiv",
    primaryClass = "gr-qc",
    doi = "10.1103/PhysRevD.111.104019",
    journal = "Phys. Rev. D",
    volume = "111",
    number = "10",
    pages = "104019",
    year = "2025"
}

@article{Fairhurst:2019vut,
    author = "Fairhurst, Stephen and Green, Rhys and Hoy, Charlie and Hannam, Mark and Muir, Alistair",
    title = "{Two-harmonic approximation for gravitational waveforms from precessing binaries}",
    eprint = "1908.05707",
    archivePrefix = "arXiv",
    primaryClass = "gr-qc",
    reportNumber = "LIGO-P1900225",
    doi = "10.1103/PhysRevD.102.024055",
    journal = "Phys. Rev. D",
    volume = "102",
    number = "2",
    pages = "024055",
    year = "2020"
}

@article{Yunes:2009yz,
    author = "Yunes, Nicolas and Arun, K. G. and Berti, Emanuele and Will, Clifford M.",
    title = "{Post-Circular Expansion of Eccentric Binary Inspirals: Fourier-Domain Waveforms in the Stationary Phase Approximation}",
    eprint = "0906.0313",
    archivePrefix = "arXiv",
    primaryClass = "gr-qc",
    doi = "10.1103/PhysRevD.80.084001",
    journal = "Phys. Rev. D",
    volume = "80",
    number = "8",
    pages = "084001",
    year = "2009",
    note = "[Erratum: Phys.Rev.D 89, 109901 (2014)]"
}

@article{Tanay:2016zog,
    author = "Tanay, Sashwat and Haney, Maria and Gopakumar, Achamveedu",
    title = "{Frequency and time domain inspiral templates for comparable mass compact binaries in eccentric orbits}",
    eprint = "1602.03081",
    archivePrefix = "arXiv",
    primaryClass = "gr-qc",
    doi = "10.1103/PhysRevD.93.064031",
    journal = "Phys. Rev. D",
    volume = "93",
    number = "6",
    pages = "064031",
    year = "2016"
}

@article{Bose:2021pcw,
    author = "Bose, Nirban and Pai, Archana",
    title = "{Effective chirp mass in the inspiral frequency evolution of the nonspinning eccentric compact binary}",
    eprint = "2107.14736",
    archivePrefix = "arXiv",
    primaryClass = "gr-qc",
    reportNumber = "LIGO-P2100280",
    doi = "10.1103/PhysRevD.104.124021",
    journal = "Phys. Rev. D",
    volume = "104",
    number = "12",
    pages = "124021",
    year = "2021"
}

@article{KAGRA:2025oiz,
    author = "Abac, A. G. and others",
    collaboration = "KAGRA, Virgo, LIGO Scientific",
    title = "{GW250114: Testing Hawking{\textquoteright}s Area Law and the Kerr Nature of Black Holes}",
    eprint = "2509.08054",
    archivePrefix = "arXiv",
    primaryClass = "gr-qc",
    reportNumber = "LIGO-P2500421",
    doi = "10.1103/kw5g-d732",
    journal = "Phys. Rev. Lett.",
    volume = "135",
    number = "11",
    pages = "111403",
    year = "2025"
}

@article{Romero-Shaw:2022fbf,
    author = "Romero-Shaw, Isobel M. and Gerosa, Davide and Loutrel, Nicholas",
    title = "{Eccentricity or spin precession? Distinguishing subdominant effects in gravitational-wave data}",
    eprint = "2211.07528",
    archivePrefix = "arXiv",
    primaryClass = "astro-ph.HE",
    doi = "10.1093/mnras/stad031",
    journal = "Mon. Not. Roy. Astron. Soc.",
    volume = "519",
    number = "4",
    pages = "5352--5357",
    year = "2023"
}

@unpublished{Divyajyoti:2025cwq,
    author = "Divyajyoti and others",
    title = "{Biased parameter inference of eccentric, spin-precessing binary black holes}",
    eprint = "2510.04332",
    archivePrefix = "arXiv",
    primaryClass = "gr-qc",
    month = "10",
    year = "2025"
}

@article{KAGRA:2013rdx,
    author = "Abbott, B. P. and others",
    collaboration = "KAGRA, LIGO Scientific, Virgo",
    title = "{Prospects for observing and localizing gravitational-wave transients with Advanced LIGO, Advanced Virgo and KAGRA}",
    eprint = "1304.0670",
    archivePrefix = "arXiv",
    primaryClass = "gr-qc",
    reportNumber = "LIGO-P1200087, VIR-0288A-12, JGW-P1808427",
    doi = "10.1007/s41114-020-00026-9",
    journal = "Living Rev. Rel.",
    volume = "19",
    pages = "1",
    year = "2016"
}

@article{LIGOScientific:2014pky,
    author = "Aasi, J. and others",
    collaboration = "LIGO Scientific",
    title = "{Advanced LIGO}",
    eprint = "1411.4547",
    archivePrefix = "arXiv",
    primaryClass = "gr-qc",
    doi = "10.1088/0264-9381/32/7/074001",
    journal = "Class. Quant. Grav.",
    volume = "32",
    pages = "074001",
    year = "2015"
}

@article{VIRGO:2014yos,
    author = "Acernese, F. and others",
    collaboration = "VIRGO",
    title = "{Advanced Virgo: a second-generation interferometric gravitational wave detector}",
    eprint = "1408.3978",
    archivePrefix = "arXiv",
    primaryClass = "gr-qc",
    doi = "10.1088/0264-9381/32/2/024001",
    journal = "Class. Quant. Grav.",
    volume = "32",
    number = "2",
    pages = "024001",
    year = "2015"
}

@article{KAGRA:2020tym,
    author = "Akutsu, T. and others",
    collaboration = "KAGRA",
    title = "{Overview of KAGRA: Detector design and construction history}",
    eprint = "2005.05574",
    archivePrefix = "arXiv",
    primaryClass = "physics.ins-det",
    doi = "10.1093/ptep/ptaa125",
    journal = "PTEP",
    volume = "2021",
    number = "5",
    pages = "05A101",
    year = "2021"
}

@article{Somiya:2011np,
    author = "Somiya, Kentaro",
    editor = "Hannam, Mark and Sutton, Patrick and Hild, Stefan and van den Broeck, Chris",
    collaboration = "KAGRA",
    title = "{Detector configuration of KAGRA: The Japanese cryogenic gravitational-wave detector}",
    eprint = "1111.7185",
    archivePrefix = "arXiv",
    primaryClass = "gr-qc",
    doi = "10.1088/0264-9381/29/12/124007",
    journal = "Class. Quant. Grav.",
    volume = "29",
    pages = "124007",
    year = "2012"
}

@article{LIGOScientific:2025brd,
    author = "Abac, A. G. and others",
    collaboration = "LIGO Scientific, KAGRA",
    title = "{GW241011 and GW241110: Exploring Binary Formation and Fundamental Physics with Asymmetric, High-spin Black Hole Coalescences}",
    eprint = "2510.26931",
    archivePrefix = "arXiv",
    primaryClass = "astro-ph.HE",
    reportNumber = "LIGO-P2500402",
    doi = "10.3847/2041-8213/ae0d54",
    journal = "Astrophys. J. Lett.",
    volume = "993",
    number = "1",
    pages = "L21",
    year = "2025"
}

@article{LIGOScientific:2025rid,
    author = "Abac, A. G. and others",
    collaboration = "LIGO Scientific, Virgo, KAGRA",
    title = "{GW250114: Testing Hawking{\textquoteright}s Area Law and the Kerr Nature of Black Holes}",
    eprint = "2509.08054",
    archivePrefix = "arXiv",
    primaryClass = "gr-qc",
    reportNumber = "LIGO-P2500421",
    doi = "10.1103/kw5g-d732",
    journal = "Phys. Rev. Lett.",
    volume = "135",
    number = "11",
    pages = "111403",
    year = "2025"
}

@unpublished{LIGOScientific:2025obp,
    author = "Abac, A. G. and others",
    collaboration = "LIGO Scientific, VIRGO, KAGRA",
    title = "{Black Hole Spectroscopy and Tests of General Relativity with GW250114}",
    eprint = "2509.08099",
    archivePrefix = "arXiv",
    primaryClass = "gr-qc",
    reportNumber = "LIGO P2500461",
    month = "9",
    year = "2025"
}

@article{LIGOScientific:2025rsn,
    author = "Abac, A. G. and others",
    collaboration = "LIGO Scientific, VIRGO, KAGRA",
    title = "{GW231123: A Binary Black Hole Merger with Total Mass 190{\textendash}265 M$_{\odot}$}",
    eprint = "2507.08219",
    archivePrefix = "arXiv",
    primaryClass = "astro-ph.HE",
    reportNumber = "DCC: P2500026-v6",
    doi = "10.3847/2041-8213/ae0c9c",
    journal = "Astrophys. J. Lett.",
    volume = "993",
    number = "1",
    pages = "L25",
    year = "2025"
}

@Article{         harris2020array,
 title         = {Array programming with {NumPy}},
 author        = {Charles R. Harris and K. Jarrod Millman and St{\'{e}}fan J.
                 van der Walt and Ralf Gommers and Pauli Virtanen and David
                 Cournapeau and Eric Wieser and Julian Taylor and Sebastian
                 Berg and Nathaniel J. Smith and Robert Kern and Matti Picus
                 and Stephan Hoyer and Marten H. van Kerkwijk and Matthew
                 Brett and Allan Haldane and Jaime Fern{\'{a}}ndez del
                 R{\'{i}}o and Mark Wiebe and Pearu Peterson and Pierre
                 G{\'{e}}rard-Marchant and Kevin Sheppard and Tyler Reddy and
                 Warren Weckesser and Hameer Abbasi and Christoph Gohlke and
                 Travis E. Oliphant},
 year          = {2020},
 month         = sep,
 journal       = {Nature},
 volume        = {585},
 number        = {7825},
 pages         = {357--362},
 doi           = {10.1038/s41586-020-2649-2},
 publisher     = {Springer Science and Business Media {LLC}},
 url           = {https://doi.org/10.1038/s41586-020-2649-2}
}

@ARTICLE{2020SciPy-NMeth,
  author  = {Virtanen, Pauli and Gommers, Ralf and Oliphant, Travis E. and
            Haberland, Matt and Reddy, Tyler and Cournapeau, David and
            Burovski, Evgeni and Peterson, Pearu and Weckesser, Warren and
            Bright, Jonathan and {van der Walt}, St{\'e}fan J. and
            Brett, Matthew and Wilson, Joshua and Millman, K. Jarrod and
            Mayorov, Nikolay and Nelson, Andrew R. J. and Jones, Eric and
            Kern, Robert and Larson, Eric and Carey, C J and
            Polat, {\.I}lhan and Feng, Yu and Moore, Eric W. and
            {VanderPlas}, Jake and Laxalde, Denis and Perktold, Josef and
            Cimrman, Robert and Henriksen, Ian and Quintero, E. A. and
            Harris, Charles R. and Archibald, Anne M. and
            Ribeiro, Ant{\^o}nio H. and Pedregosa, Fabian and
            {van Mulbregt}, Paul and {SciPy 1.0 Contributors}},
  title   = {{{SciPy} 1.0: Fundamental Algorithms for Scientific
            Computing in Python}},
  journal = {Nature Methods},
  year    = {2020},
  volume  = {17},
  pages   = {261--272},
  adsurl  = {https://rdcu.be/b08Wh},
  doi     = {10.1038/s41592-019-0686-2},
}

@Article{Hunter:2007,
  Author    = {Hunter, J. D.},
  Title     = {Matplotlib: A 2D graphics environment},
  Journal   = {Computing in Science \& Engineering},
  Volume    = {9},
  Number    = {3},
  Pages     = {90--95},
  publisher = {IEEE COMPUTER SOC},
  doi       = {10.1109/MCSE.2007.55},
  year      = 2007
}

@misc{2024zndo..10940210A,
       author = {{Ashton}, Gregory and {Talbot}, Colm and {Roy}, Soumen and {Pratten}, Geraint and {Pang}, Tsun-Ho and {Agathos}, Michalis and {Baka}, Tomasz and {S{\"a}nger}, Elise and {Mehta}, Ajit and {Steinhoff}, Jan and {Maggio}, Elisa and {Ghosh}, Abhirup and {Vijaykumar}, Aditya},
        title = "{Bilby TGR}",
         year = 2024,
        month = apr,
          eid = {10.5281/zenodo.10940210},
          doi = {10.5281/zenodo.10940210},
      version = {v0.1a2},
    publisher = {Zenodo},
       adsurl = {https://ui.adsabs.harvard.edu/abs/2024zndo..10940210A}
}

@article{bilby_pipe_paper,
    author = "Romero-Shaw, I. M. and others",
    title = "{Bayesian inference for compact binary coalescences with bilby: validation and application to the first LIGO\textendash{}Virgo gravitational-wave transient catalogue}",
    eprint = "2006.00714",
    archivePrefix = "arXiv",
    primaryClass = "astro-ph.IM",
    doi = "10.1093/mnras/staa2850",
    journal = "Mon. Not. Roy. Astron. Soc.",
    volume = "499",
    number = "3",
    pages = "3295--3319",
    year = "2020"
}

@misc{sergey_koposov_2025_17268284,
  author       = {Sergey Koposov and
                  others},
  title        = {joshspeagle/dynesty: v3.0.0},
  month        = oct,
  year         = 2025,
  publisher    = {Zenodo},
  version      = {v3.0.0},
  doi          = {10.5281/zenodo.17268284},
  url          = {https://doi.org/10.5281/zenodo.17268284},
  swhid        = {swh:1:dir:b3340f6aaa931bae6c6bcff6e7ddeb89ca508a15
                   ;origin=https://doi.org/10.5281/zenodo.3348367;vis
                   it=swh:1:snp:fa92422c27d5611b107216e9b766a871bb43b
                   bee;anchor=swh:1:rel:94af3a8ef6e50d997258e152c3833
                   3550bd7463f;path=joshspeagle-dynesty-217bc94
                  },
}

@article{Hoy:2020vys,
    author = "Hoy, Charlie and Raymond, Vivien",
    title = "{PESummary: the code agnostic Parameter Estimation Summary page builder}",
    eprint = "2006.06639",
    archivePrefix = "arXiv",
    primaryClass = "astro-ph.IM",
    reportNumber = "LIGO-P2000156",
    doi = "10.1016/j.softx.2021.100765",
    journal = "SoftwareX",
    volume = "15",
    pages = "100765",
    year = "2021"
}

@article{Harrower01062003,
author = {Mark Harrower and Cynthia A. Brewer},
title = {ColorBrewer.org: An Online Tool for Selecting Colour Schemes for Maps},
journal = {The Cartographic Journal},
volume = {40},
number = {1},
pages = {27--37},
year = {2003},
publisher = {Taylor \& Francis},
doi = {10.1179/000870403235002042},
URL = { 
https://www.tandfonline.com/doi/abs/10.1179/000870403235002042
},
eprint = { 
https://www.tandfonline.com/doi/pdf/10.1179/000870403235002042
},
}

@unpublished{Chandra:2025gw250114,
  title        = {From Source Properties to Strong-Field Tests: A Multipronged Analysis of GW250114 with an Effective One-Body Model for Generic Orbits},
  author       = {Chandra, Koustav and Gamba, Rossella and Chiaramello, Danilo},
  note         = {in preparation},
  year         = {2025}
}

@article{Gamba:2024cvy,
    author = "Gamba, Rossella and Chiaramello, Danilo and Neogi, Sayan",
    title = "{Toward efficient effective-one-body models for generic, nonplanar orbits}",
    eprint = "2404.15408",
    archivePrefix = "arXiv",
    primaryClass = "gr-qc",
    doi = "10.1103/PhysRevD.110.024031",
    journal = "Phys. Rev. D",
    volume = "110",
    number = "2",
    pages = "024031",
    year = "2024"
}

@article{Nagar:2024oyk,
    author = "Nagar, Alessandro and Chiaramello, Danilo and Gamba, Rossella and Albanesi, Simone and Bernuzzi, Sebastiano and Fantini, Veronica and Panzeri, Mattia and Rettegno, Piero",
    title = "{Effective-one-body waveform model for noncircularized, planar, coalescing black hole binaries. II. High accuracy by improving logarithmic terms in resummations}",
    eprint = "2407.04762",
    archivePrefix = "arXiv",
    primaryClass = "gr-qc",
    doi = "10.1103/PhysRevD.111.064050",
    journal = "Phys. Rev. D",
    volume = "111",
    number = "6",
    pages = "064050",
    year = "2025"
}

@unpublished{Albanesi:2025txj,
    author = "Albanesi, Simone and Gamba, Rossella and Bernuzzi, Sebastiano and Fontbut{\'e}, Joan and Gonzalez, Alejandra and Nagar, Alessandro",
    title = "{Effective-one-body modeling for generic compact binaries with arbitrary orbits}",
    eprint = "2503.14580",
    archivePrefix = "arXiv",
    primaryClass = "gr-qc",
    month = "3",
    year = "2025"
}

@article{Capote:2024rmo,
    author = "Capote, E. and others",
    title = "{Advanced LIGO detector performance in the fourth observing run}",
    eprint = "2411.14607",
    archivePrefix = "arXiv",
    primaryClass = "gr-qc",
    reportNumber = "LIGO-P2400256",
    doi = "10.1103/PhysRevD.111.062002",
    journal = "Phys. Rev. D",
    volume = "111",
    number = "6",
    pages = "062002",
    year = "2025"
}

@article{aLIGO:2020wna,
    author = "Buikema, Aaron and others",
    collaboration = "aLIGO",
    title = "{Sensitivity and performance of the Advanced LIGO detectors in the third observing run}",
    eprint = "2008.01301",
    archivePrefix = "arXiv",
    primaryClass = "astro-ph.IM",
    doi = "10.1103/PhysRevD.102.062003",
    journal = "Phys. Rev. D",
    volume = "102",
    number = "6",
    pages = "062003",
    year = "2020"
}

@article{Virgo:2019juy,
    author = "Acernese, F. and others",
    collaboration = "Virgo",
    title = "{Increasing the Astrophysical Reach of the Advanced Virgo Detector via the Application of Squeezed Vacuum States of Light}",
    doi = "10.1103/PhysRevLett.123.231108",
    journal = "Phys. Rev. Lett.",
    volume = "123",
    number = "23",
    pages = "231108",
    year = "2019"
}

@unpublished{LIGOScientific:2026qni,
    author = "Abac, A. G. and others",
    collaboration = "LIGO Scientific, VIRGO, KAGRA",
    title = "{GWTC-4.0: Tests of General Relativity. I. Overview and General Tests}",
    eprint = "2603.19019",
    archivePrefix = "arXiv",
    primaryClass = "gr-qc",
    reportNumber = "LIGO-P2500065",
    month = "3",
    year = "2026"
}

@unpublished{LIGOScientific:2026fcf,
    author = "Abac, A. G. and others",
    collaboration = "LIGO Scientific, VIRGO, KAGRA",
    title = "{GWTC-4.0: Tests of General Relativity. II. Parameterized Tests}",
    eprint = "2603.19020",
    archivePrefix = "arXiv",
    primaryClass = "gr-qc",
    reportNumber = "LIGO-P2500066",
    month = "3",
    year = "2026"
}

@unpublished{LIGOScientific:2026wpt,
    author = "Abac, A. G. and others",
    collaboration = "LIGO Scientific, VIRGO, KAGRA",
    title = "{GWTC-4.0: Tests of General Relativity. III. Tests of the Remnants}",
    eprint = "2603.19021",
    archivePrefix = "arXiv",
    primaryClass = "gr-qc",
    reportNumber = "LIGO-P2500067",
    month = "3",
    year = "2026"
}

@article{Mehta:2022pcn,
    author = "Mehta, Ajit Kumar and Buonanno, Alessandra and Cotesta, Roberto and Ghosh, Abhirup and Sennett, Noah and Steinhoff, Jan",
    title = "{Tests of general relativity with gravitational-wave observations using a flexible theory-independent method}",
    eprint = "2203.13937",
    archivePrefix = "arXiv",
    primaryClass = "gr-qc",
    reportNumber = "LIGO-P2200083-v2",
    doi = "10.1103/PhysRevD.107.044020",
    journal = "Phys. Rev. D",
    volume = "107",
    number = "4",
    pages = "044020",
    year = "2023"
}

@article{Saini:2022igm,
    author = "Saini, Pankaj and Favata, Marc and Arun, K. G.",
    title = "{Systematic bias on parametrized tests of general relativity due to neglect of orbital eccentricity}",
    eprint = "2203.04634",
    archivePrefix = "arXiv",
    primaryClass = "gr-qc",
    reportNumber = "LIGO Preprint No. P2200073",
    doi = "10.1103/PhysRevD.106.084031",
    journal = "Phys. Rev. D",
    volume = "106",
    number = "8",
    pages = "084031",
    year = "2022"
}

@article{Chandramouli:2024vhw,
    author = "Chandramouli, Rohit S. and Prokup, Kaitlyn and Berti, Emanuele and Yunes, Nicol{\'a}s",
    title = "{Systematic biases due to waveform mismodeling in parametrized post-Einsteinian tests of general relativity: The impact of neglecting spin precession and higher modes}",
    eprint = "2410.06254",
    archivePrefix = "arXiv",
    primaryClass = "gr-qc",
    doi = "10.1103/PhysRevD.111.044026",
    journal = "Phys. Rev. D",
    volume = "111",
    number = "4",
    pages = "044026",
    year = "2025"
}

@article{Mezzasoma:2022pjb,
    author = "Mezzasoma, Simone and Yunes, Nicol{\'a}s",
    title = "{Theory-agnostic framework for inspiral tests of general relativity with higher-harmonic gravitational waves}",
    eprint = "2203.15934",
    archivePrefix = "arXiv",
    primaryClass = "gr-qc",
    doi = "10.1103/PhysRevD.106.024026",
    journal = "Phys. Rev. D",
    volume = "106",
    number = "2",
    pages = "024026",
    year = "2022"
}

@article{Pitte:2023ltw,
    author = "Pitte, Chantal and Baghi, Quentin and Marsat, Sylvain and Besan{\c{c}}on, Marc and Petiteau, Antoine",
    title = "{Detectability of higher harmonics with LISA}",
    eprint = "2304.03142",
    archivePrefix = "arXiv",
    primaryClass = "gr-qc",
    doi = "10.1103/PhysRevD.108.044053",
    journal = "Phys. Rev. D",
    volume = "108",
    number = "4",
    pages = "044053",
    year = "2023"
}

@article{Yi:2025pxe,
    author = "Yi, Sophia and Iacovelli, Francesco and Marsat, Sylvain and Wadekar, Digvijay and Berti, Emanuele",
    title = "{Systematic biases in parameter estimation on LISA binaries: The effect of excluding higher harmonics for non-spinning binaries}",
    eprint = "2502.12237",
    archivePrefix = "arXiv",
    primaryClass = "gr-qc",
    doi = "10.1103/3gs3-gmb4",
    journal = "Phys. Rev. D",
    volume = "112",
    number = "12",
    pages = "124063",
    year = "2025"
}

@unpublished{Yi:2026ucv,
    author = "Yi, Sophia and Iacovelli, Francesco and Berti, Emanuele and Chandramouli, Rohit S. and Marsat, Sylvain and Wadekar, Digvijay and Yunes, Nicol{\'a}s",
    title = "{Systematic biases in parameter estimation on LISA binaries. II. The effect of excluding higher harmonics for spin-aligned, high-mass binaries}",
    eprint = "2602.09088",
    archivePrefix = "arXiv",
    primaryClass = "gr-qc",
    month = "2",
    year = "2026"
}
\end{document}